# Engineering Dion-Jacobson Perovskites in Polariton Waveguides

A. Coriolano[1,2], A. Moliterni*[3], F. Todisco[1], L. Polimeno[1], R. Mastria[1], V. Olieric[4], C. Giacobbe[5], M. De Giorgi[1], D. Ballarini[1], A. Rizzo[1], G. Gigli[1,2], C. Giannini[3], I. Viola[6], D. Sanvitto*[1] and L. De Marco*[1]

[1] *CNR NANOTEC, Institute of Nanotechnology, Via Monteroni, Lecce 73100, Italy*

[2] *Dipartimento di Matematica e Fisica E. De Giorgi, Università del Salento, Campus Ecotekne, Via Monteroni, Lecce 73100, Italy*

[3]*CNR, Institute of Crystallography, Via Amendola, 122/O, Bari 70126, Italy*

[4]*Swiss Light Source, Paul Scherrer Institute, 5232 Villigen PSI, Switzerland*

[5]*European Synchrotron Radiation Facility, Avenue des Martyrs 71, 38040 Grenoble, France*

[6] *CNR NANOTEC, Institute of Nanotechnology, Soft and Living Matter Laboratory, c/o Dipt. di Fisica, Sapienza Università, P.le A. Moro 2, I-00185, Rome, Italy.*

## Abstract

Hybrid two-dimensional perovskites hold considerable promise as semiconductors for a wide range of optoelectronic applications. Many efforts are addressed to exploit the potential of these materials by tailoring their characteristics. In this work, the optical properties and electronic band structure in three new Dion-Jacobson (DJ) perovskites (PVKs) are engineered by modulating their structural distortion. Two different interlayer cations: 1-6, Hexamethylendiammonium, HE, and 3-(Dimethylamino)-1-propylammonium, DMPA, have been selected to investigate the role of the cation length and the ammonium binding group on the crystalline structure. This study provides new insights into the understanding of the structure-property relationship in DJ perovskites and demonstrates that exciton characteristics can be easily modulated with the judicious design of the organic cations.

DJ PVKs developed in this work were also grown as size-controlled single crystal microwires through a microfluidic-assisted synthesis technique and integrated in a nanophotonic device. The DJ PVK microwire acts as a waveguide exhibiting strong light-matter coupling between the crystal optical modes and DJ PVK exciton. Through the investigation of these polariton waveguides, the nature of the double peak emission, which is often observed in these materials and whose nature is largely debated in the literature, is demonstrated originating from the hybrid polariton state.

Keywords: 2D perovskites; Dion-Jacobson; microfluidics; waveguides; polaritons

## Introduction

Hybrid organic – inorganic metal halide perovskites (PVKs) have radically transformed the field of optoelectronics thanks to their unique photophysical and electrical properties, combined with their ease of synthesis and manipulation.[1–3] Two-dimensional (2D) PVKs, consisting of alternating layers of metal-halide octahedra and bulky organic cations, have attracted great interest due to their natural multiple quantum wells (MQW) structure, which provides these materials with appealing and tailorable optical properties observed at room temperature (RT), including a high oscillator strength and narrow photoluminescence.[4–6] In addition, the high refractive index of 2D PVKs,[7] combined with their high optical non-linearities at RT,[8] make them extremely intriguing for polariton systems that exploit the strong coupling between exciton and optical modes.[9–13]

The structural configuration of 2D PVKs offers great versatility to master their properties, for example, the different halide elements (*i.e.*, Cl, Br, and I) of the metal-halide octahedra enable the tuning of the band gap over a wide energy range,[14–16] while the bulky organic cations offer a full set of possibilities to induce opportune modification in the photophysics of 2D PVKs, as demonstrated in recent reports.[17–19] The use of long-chain organic molecules, such as dodecylammonium, increases the bandgap of 2D perovskites compared to shorter molecules, with consequent blue-shift of the photoluminescence.[20–22] The packing arrangements adopted by organic ligands can be exploited to deform the inorganic slabs, inducing both intra- and inter-octahedral distortions with strong exciton–lattice coupling, influencing emission color[14] and enabling broadband and white emission at room temperature[23–26] with great potential in white light-emitting diodes.

The choice of the organic component also provides a way to control the tuning of exciton confinement in terms of orientation, vibrational properties, and polarization of the transition dipoles[9,27] or favoring the birefringence of the material.[13] In addition, the mobility of free carriers can be enhanced by selecting organic cations based on amines with high dielectric constant, ε, such as ethanolamine or ethylenediamine instead of the typical butylamine or phenethylamine, which have low ε.[28]

These intriguing properties of 2D PVK are combined with low-cost processing, improved stability, and endurance to moisture, making them more robust under environmental conditions, which is one of the most critical factor for the performance and lifetime of optoelectronic devices.[29,30]

The most extensively studied 2D PVKs are in the Ruddlesden–Popper (RP) phase, with formula $(A)_2MX_4$ where A is a monovalent organic ammonium cation (such as butylammonium, phenethylammonium, etc.), M is the divalent metal cation (such as Pb, Sn, Ge) and X are the halide anions (such as Br, I, Cl).

However, 2D PVKs with Dion-Jacobson (DJ) phase have been recently introduced[31] aiming at the improvement of performances of 2D-PVKs based optoelectronic devices. DJ PVKs have formula $A'MX_4$ where A' is a divalent organic ammonium cation such as 3-(aminomethyl)piperidinium, 1,3-propanediammonium or 1,4-phenylenedimethanammonium. Compared with RP PVKs, which are characterized by pairs of interdigitated monovalent organic cations per unit cell, DJ PVKs have divalent spacers, requiring only one cation per unit cell. This results in a tighter connection between the inorganic sheets enabled by strong hydrogen bonds formed between diammonium cations and octahedra instead of weak van der Waals interactions present in the RP phase.[32–34] In addition, the use of organic dications results in the shortening of the interlayer distance, thus promoting carriers hopping/tunneling between different layers.[35,36] Owing to this, DJ PVKs are considered a great opportunity to improve performance of optoelectronic devices in terms of efficiency and stability.[37–42] However, DJ PVKs possess much more potential than just enhanced stability and improved charge transport. Their unique structure, characterized by the strong interaction between the organic and the inorganic layers, allows for a wide range of tunability in their optoelectronic properties, which provides DJ PVKs with an unprecedented versatile character.

Despite this and the rich variety of diamines available, only a few of them have been successfully employed in DJ PVKs, leaving room for further expanding the hybrid semiconductor library toward a better understanding of structure-property relationships in this class of materials.

Herein we expand the library of DJ PVK demonstrating the synthesis of three DJ perovskites crystals, namely $HEPbI_4$ and two polymorphs of $DMPAPbI_4$ (HE= 1-6, Hexamethylendiamine; DMPA= 3-(Dimethylamino)-1-propylamine). The different organic cations have been wisely selected to understand how the molecular structure of the organic dication affects DJ PVK photophysics. The crystal structure of $HEPbI_4$ and $DMPAPbI_4$ has been finely analyzed by synchrotron single-crystal X-Ray Diffraction (SCXRD). Besides, micro-photoluminescence (micro-PL) and spatial-resolved photoluminescence (SR-PL) measurements were used to study their photophysics. In this way, we were able to correlate the optical properties with the structural characteristics of these engineered DJ PVKs single crystals.

Furthermore, we revealed the presence of an additional PL peak at lower energies than the main excitonic transition, which is detectable only at the edge of the crystals. The nature of this peak is still debated in literature.[22,43–45] To provide an explanation of this phenomenon, we synthesized crystalline DJ PVK microwires through a microfluidic-assisted confined crystal growth[46,47] aiming to obtain high-quality, size- and shape-controlled crystals. We integrated these crystals with opportune optical gratings, and performed an investigation in strong coupling regime. Our results demonstrate unambiguously that such anomalous emission is related to polariton modes arising from strong light-matter interactions between waveguided modes confined in the PVK crystals and DJ PVK excitons.

**Results and discussion**

2D DJ perovskites have been prepared by using two different diamines, namely 1-6, Hexamethylendiamine, HE, and 3-(Dimethylamino)-1-propylamine, DMPA, as organic dications. HE and DMPA have been selected to investigate the effect of the cation length and the type of ammonium binding group on the corresponding perovskite structure. HE has symmetric molecular structure with a hexyl chain (C6) and two terminal primary amines as binding groups (Figure 1a, left) while DMPA has shorter propyl chain (C3) and an asymmetric molecular structure carrying a primary amine on one head and a dimethyl-substituted tertiary amine at the other extremity (Figure 1a, right). High quality perovskites single crystals have been synthesized by a solution-based growth method which exploit the slow cooling of the precursor solution to induce a controlled crystallization.[23,24] The synthesis involves the use of hydroiodic acid (HI) which served as both solvent and iodide source. In detail, the metal source lead oxide (PbO) was dissolved in HI under heating and stirring to obtain a clear solution. Diamine, either HE or DMPA, was then added and perovskite crystals formed by slow cooling to room temperature. A sketch of the synthetic process is shown in Figure S1. Figure S2 shows scanning electron microscope (SEM) images of the as obtained micron-sized $HEPbI_4$ single crystals on glass substrate.

Typical images of $HEPbI_4$ and two polymorphic $DMPAPbI_4$ (from this point onwards called only HE and DMPA for simplicity) perovskite bulk single crystals on glass substrate are reported in Figure 1b-d.

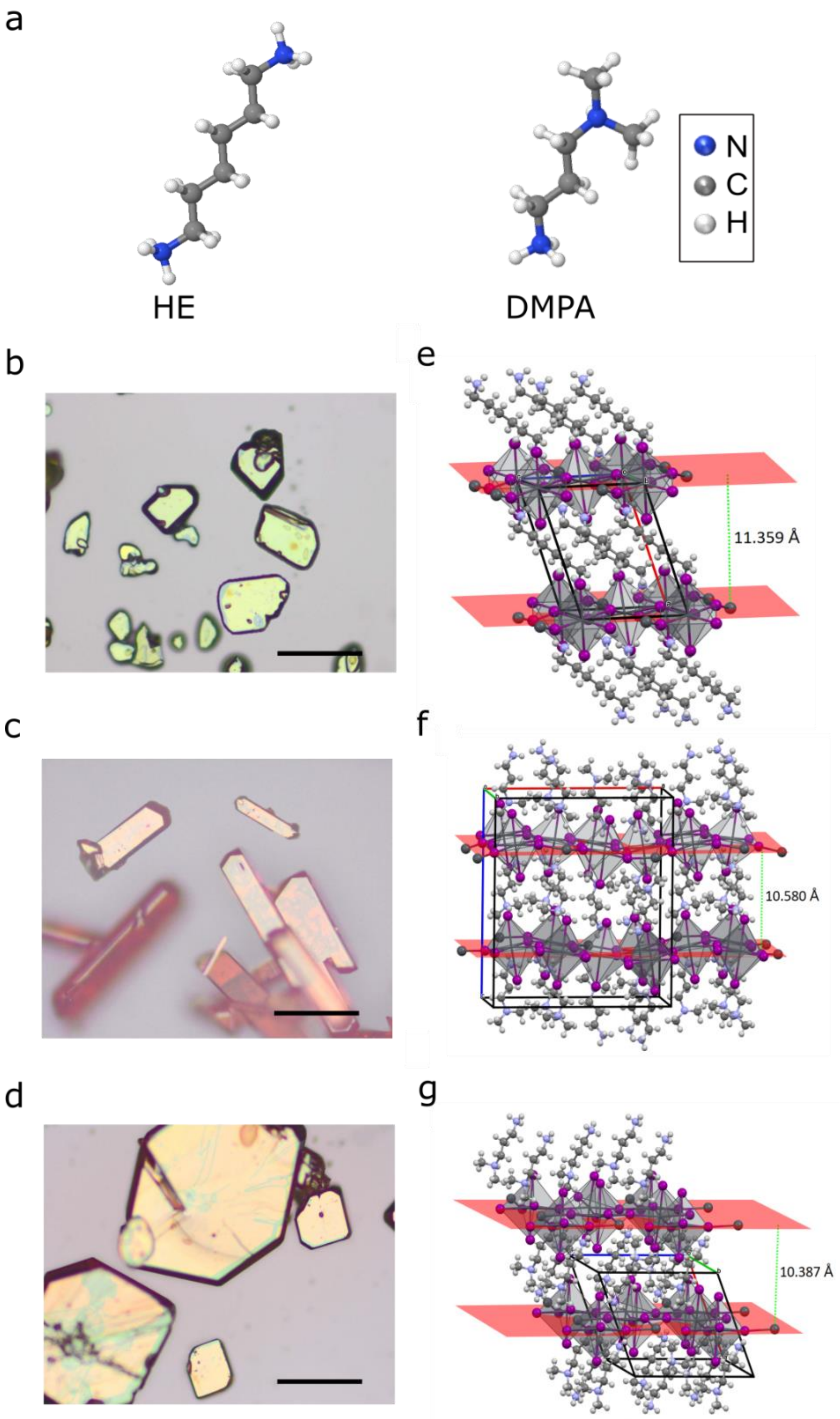

**Figure 1.** (a) Molecular structures of the diamines explored in this paper: 1,6-Hexamethylendiamine (left) and 3-(Dimethylamino)-1-propylamine) (right). (b-d) Optical image of $HEPbI_4$ single crystals (b) and $DMPAPbI_4$ single crystals phase 1 (c) and phase 2 (d). Scale bars are 100 µm. A view of the crystal packing of the investigated perovskites, showing the distance between the two nearest least-squares planes through the equatorial atoms of the $PbI_6$ octahedra for HE (e) DMPA (1) (f), DMPA (2) (g).

Interestingly, with DMPA cation we first observed the formation of reddish and elongated crystals, which we label DMPA (1), see Figure 1c. Then, when the as obtained crystals are left in the mother liquor for 24 hours at room temperature, they spontaneously evolve into more stable square-shaped light red crystals, hereafter referred to as DMPA (2), shown in Figure 1d. Phase 1 is formed by a fast crystallization kinetics which occurs in a range of time of about 1h following the cooling of the precursor solution at room temperature, while phase 2 is formed by a slower molecular packing rearrangement over time and a consequent different crystallization of the perovskite.

The crystal structures of the synthesized perovskites were determined by single crystal X-ray diffraction. HE, DMPA (1) and DMPA (2) crystallize in centrosymmetric space groups, *i.e.*, $P2_1/c$ (HE and DMPA (2)) and *Pbca* (DMPA (1), a polymorph of DMPA (2)), see Table 1.

Main experimental and structure refinement details are given in Supporting Information (Figure S3-S8 and Table S1-S7). The crystal packing of HE, DMPA (2) and DMPA (1) consists of anionic inorganic layers of corner-sharing $PbI_6$ octahedra, sandwiched between diammonium cations and usually oriented along the longest axis,[33,48] e.g., the *a* axis in the case of HE and *c* axis in the case of DMPA (1). The distance between the two nearest least squares planes through the equatorial atoms of the $PbI_6$ octahedra for HE, DMPA (1) and DMPA (2) is about 11.359 Å, 10.580 Å and 10.387 Å, respectively (see Figure 1e-g); for DMPA (1) and DMPA (2) the interplanar distances are slightly smaller than that one in HE, probably due to the longer length of the organic chain of HE and the smaller number of the H-bonds (*i.e.*, four H-bonds for HE and eleven for DMPA (2) and DMPA (1), see Tab. S3, S5 and S7, respectively).

The interactions between the polar $NH_3$ groups of the organic cations and the halogen atoms of the octahedra play a fundamental role in the crystal packing of the perovskite-derivative structures[48] and are responsible for a coordinated structural distortion.[5] The bridging equatorial distortion $Pb_{eq}$–I–$Pb_{eq}$ angle for HE is 148.63°, far from 180.0° (*i.e.*, the typical value of undistorted heavy atoms chains), and corresponds to an in-plane rotation of adjacent octahedra, similar the rotation angles observed in literature.[48] The octahedral distortion is strictly correlated to the arrangement of the organic molecule inside the perovskite structure, resulting in a different $Pb_{eq}$–I–$Pb_{eq}$ angle for both DMPA compounds where the equatorial distortion $Pb_{eq}$–I–$Pb_{eq}$ angle is 156.83° and 163.13° for DMPA (2) and DMPA (1), respectively.

| *Crystal data* | *HE* | *DMPA (2)* | *DMPA (1)* |
|---|---|---|---|
| Chemical formula | $C_6H_{18}N_2PbI_4$ | $C_5H_{16}N_2PbI_4$ | $C_5H_{16}N_2PbI_4$ |
| $M_r$ | 833.01 | 818.99 | 818.99 |
| Crystal system | Monoclinic | Monoclinic | Orthorhombic |
| Space group | $P2_1/c$ | $P2_1/c$ | *Pbca* |
| Temperature (K) | 293 | 293 | 293 |
| *a*, *b*, *c* (Å) | 11.8712(4), 8.4911(2), 9.0503(3) | 11.2176(4), 12.5073(3), 12.8819(3) | 18.406(1), 8.650(2), 20.726(1) |
| β (°) | 106.856(3) | 112.929 (3) | 90.0 |
| $V$ (Å$^3$) | 873.07(5) | 1664.55 (9) | 3299.8 (8) |
| $Z$ | 2 | 4 | 8 |

**Table 1.** Crystal data for HE, DMPA (2) and DMPA (1).

The level of octahedral distortion is related to the cation penetration:[49] to have insight into the influence of the organic cation penetration on the inorganic layer distortion, in the case of DMPA(1), DMPA(2) and HE we calculated the $NH_3$ penetration in terms of average distance of the nitrogen atoms, belonging to the terminal $NH_3$ group, from the nearest least-squares planes through the upper and lower axial iodine atoms of the $PbI_6$ octahedra (*i.e.*, $<d_N>_{U\&L}$ values in Table S8 and Figure S9-S13). In accordance with the literature results[49], the in-plane angle $Pb_{eq}$–I–$Pb_{eq}$ increased as the penetration of $NH_3$ decreased, *i.e*, as the distance between $NH_3$ and octahedra increased, (see Table S8) and, consequently, the distortion effects decreased. The type of interlayer dication in DJ perovskites directly affects the hydrogen bonds with the octahedra and thus the $Pb_{eq}$–I–$Pb_{eq}$ bond angles, determining the structural distortion and physical properties of the inorganic framework.[50] We then investigated the correlation existing between the $Pb_{eq}$–I–$Pb_{eq}$ angle values and the photoluminescence of the perovskite crystals synthesized in this work (see Figure 2). To better frame the behaviour of HE and DMPA perovskites, the results were compared with a well-known RP perovskite, the 4-fluoro-phenethylammonium iodide-based perovskite, PEAI-F, which has $Pb_{eq}$–I–$Pb_{eq}$ angle equal to 151.8°.[13,51]

We observed that as the value of the $Pb_{eq}$–I–$Pb_{eq}$ decreases, diverging from the ideal value of 180°, the photoluminescence peak undergoes a blueshift. In particular, for HE single crystals the PL peak is centred at 2.5 eV (Figure 2e, blue line), for PEAI-F the PL peak is centred at 2.34 eV (Figure 2e, orange line), while for DMPA (2) and DMPA (1) the PL peak is centred at 2.3 eV and at 2.21 eV, respectively (Figure 2e, red and black lines).

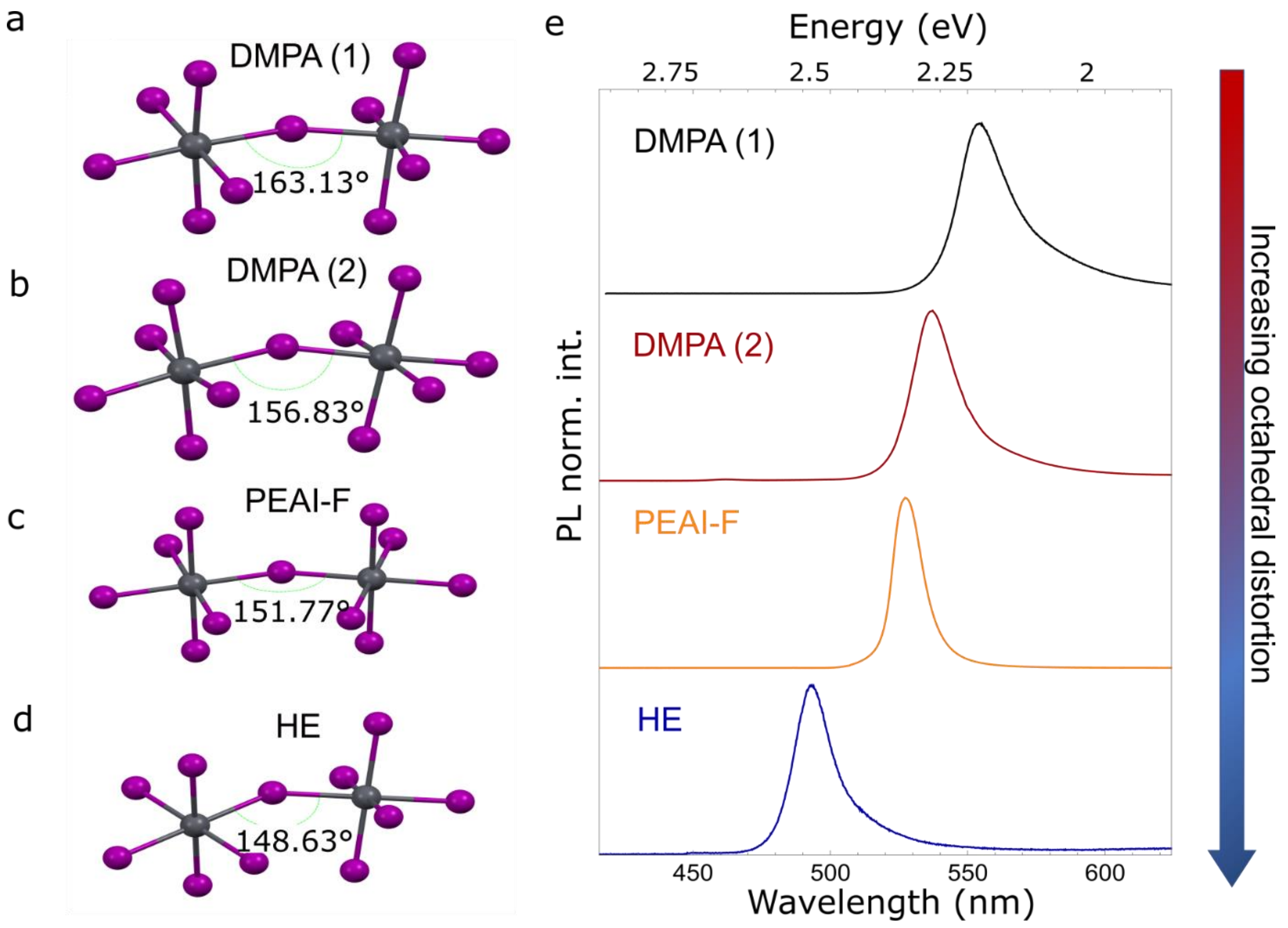

**Figure 2.** Values of the bridging in-plane distortion $Pb_{eq}$–I–$Pb_{eq}$ angle: 163.13° for DMPA (1) (a); 156.83° for DMPA (2) (b); 151.77° for PEAI-F (c); 148.63° for HE (d). (e)Photoluminescence spectra of DMPA (1) (Black line), DMPA (2) (Red line), PEAI-F (Orange line) and HE (Blue line); The spectra are vertically spaced for clarity.

This behaviour can be related to a reduced Pb-I orbital overlap upon distortion, leading to an increased splitting of the energy levels, with consequent increasing of the band gap and PL blueshift.[24,52,53] This opens the possibility to fine tune the energy gap of DJ PVKs by controlling the octahedral tilting.

To further investigate the optical features of the synthesized DJ PVKs bulk crystals, we performed the temperature-dependent PL measurements of HE and the stable DMPA (2) phase PVK crystals from 77K to RT. By decreasing temperature, for both the synthesised PVKs we observed a blueshift of the excitonic transition peak (Figure S15a and Figure S15b). The observed variation in the energy gap with the temperature can be ascribed to the combination of two main mechanisms[43,54,55] that induce an energy shift of conduction and valence bands:

(i) the temperature-dependent lattice dilatation and (ii) a temperature-dependent electron – phonon interaction. A detailed analysis of temperature dependence of the excitonic transitions in HE and DMPA (2) perovskites is reported in Supporting Information.

An additional feature of millimeter-sized 2D-PVK single-crystals reported in literature,[22,43–45] consists in the observation of an anomalous PL peak at the crystals edges. In order to shed light on this feature we performed spatial-resolved photoluminescence collecting the PL signal at different points of the crystal. By exciting and collecting the signal at the centre of the crystal (under the spot laser, inset Figure 3a and 3b, blue and red triangles, respectively), only the main excitonic peak ($E_{ex}$) for both HE and DMPA (2) perovskite (Figure 3a and 3b, continuous line) is observed. On the other hand, by exciting the crystals at the centre (triangle) and collecting the PL signal at the edge (dashed squares in the insets of Figure 3), an additional anomalous peak ($E_2$) appears, which is redshifted with respect to the principal excitonic transition (Figure 3a and 3b, dashed line).

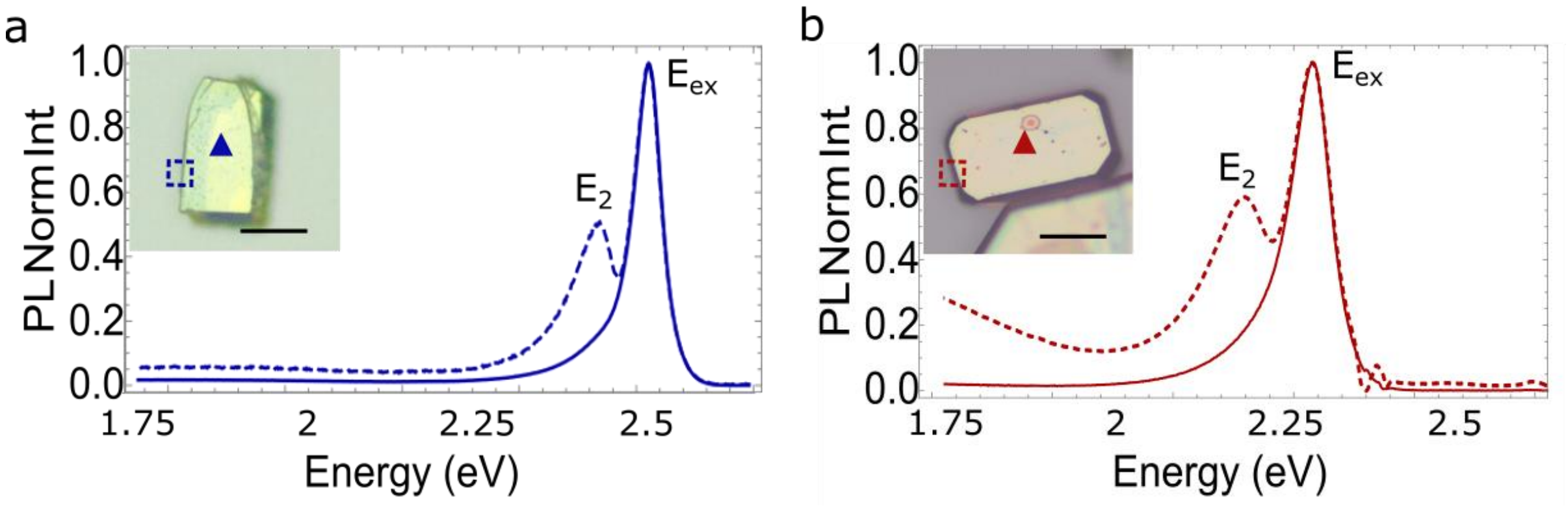

**Figure 3.** (a) Normalized PL of HE single crystal on glass substrate, exciting at the centre of the crystal and collecting the signal at the centre (Blue continuous line) and at the edge of the crystal (Blue dashed line). (Inset) Image of the measured HE crystal on glass substrate (b) Normalized PL of DMPA (2) single crystal on glass substrate, exciting at the centre of the crystal and collecting the signal at the centre (Red continuous line) and at the edge of the crystal (Red dashed line). (Inset) Image of the measured DMPA (2) crystal on glass substrate. The scale bars are 50 µm.

The nature of the lowest-energy peak is strongly debated in literature. Sheikh et al.[44] suggested a possible dual bandgap, resulting in the double PL peak: the one at lower-energy originating from the bulk, and the one at higher-energy resulting from the inorganic layers close to the surface, which show a slightly different structural distortion. In another report the additional peak detectable at the edges of the crystals was assigned to a possible electronic interaction between two neighbouring inorganic layers, probably due to a different supramolecular pathway during crystallization, which reduces the distance between the adjacent layers.[22,45] On the other hand, Wang et al.[43] argued that in a regularly shaped DJ perovskite single crystal there is an enhancement of Rashba emission through photon recycling, resulting in the presence of a lower energy peak in the PL spectrum.

Differently from previous reports, we hypothesized that the second peak detectable only at the edges of the investigated crystals, is related to a waveguide effect that occurs by total internal reflection within the PVK slab.  In order to verify this hypothesis, we thus fabricated a nanophotonic device consisting of perovskite microwire-based waveguides and a patterned polymethyl methacrylate (PMMA) grating placed over the top of the crystals designed to extract the propagating modes. Both HE and DMPA (2) microwires, with a width of 6 μm and a thickness of 300 nm, were fabricated by means of a microfluidic-assisted growth technique[46,47] and by controlling the nucleation kinetics, so as to obtain high-quality single crystals with sharp edges, and with tailored and predetermined size and shape (see Supporting Information for further details). Such a microfluidic growth technique allows, in fact, to control the evaporation rate of the precursor solution, in order to reach a controlled state of supersaturation, and thus modulate the nucleation kinetics and consequently the structure and quality of the grown crystal. Figure S16 shows a scanning electron microscope (SEM) image of microfluidic-grown single crystals, revealing a uniform and flat surface and sharp perovskite edges.

Subsequently, we fabricated a 180 nm thick 1-D PMMA grating on the top of the as grown single crystal microwires by electron beam lithography (see Figure 4a and 4b). The grating consists in an array of stripes, with a pitch of 300 nm and a filling factor (FF) of 0.5.

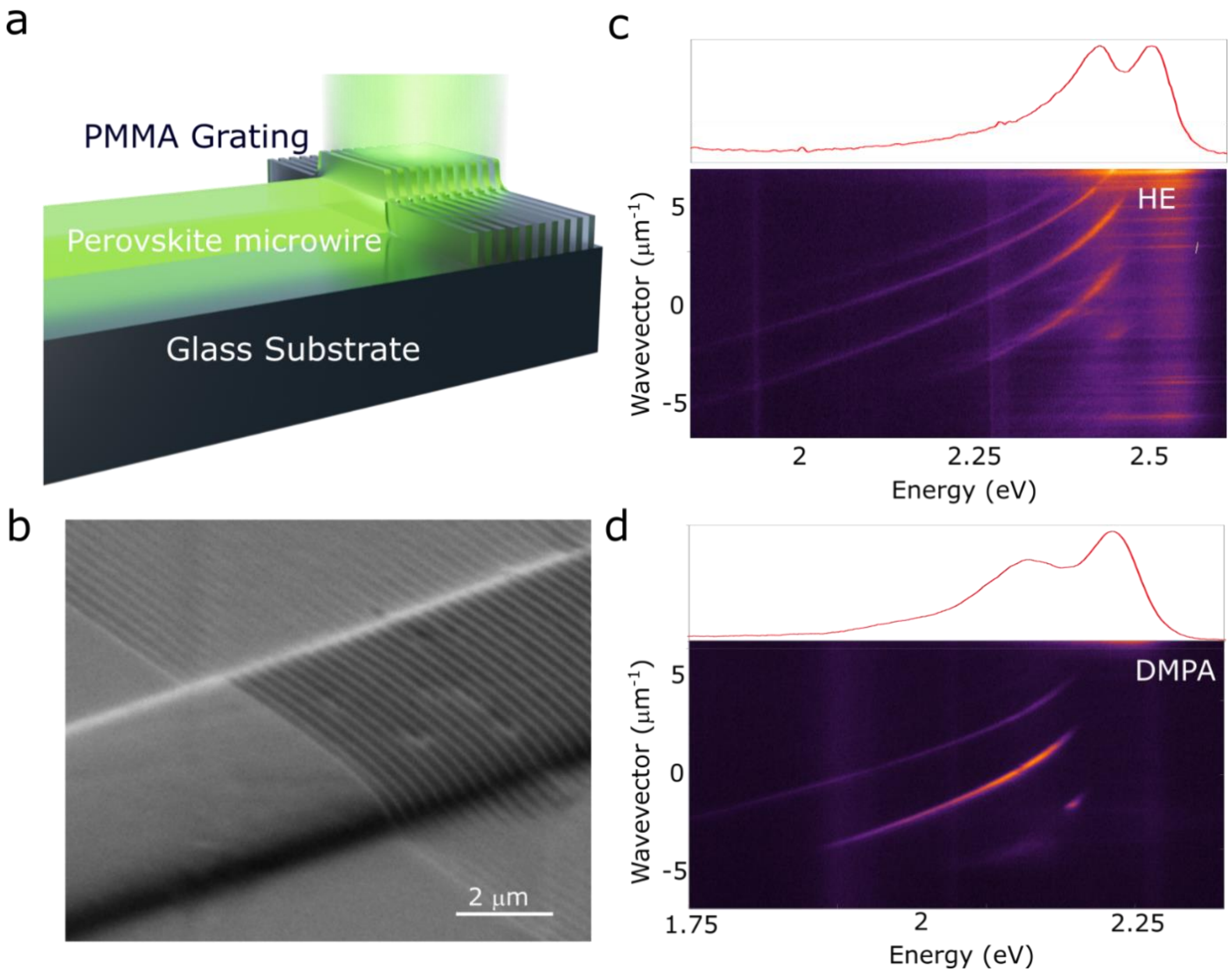

**Figure 4.** (a) Sketch of the device used in this paper composed by a perovskite microwire having a thickness of 300 nm and a width of 6 μm and a PMMA grating on top. (b) SEM image of the device. (c) Integrated PL signal for all k-vectors (Top); PL spectra plotted as energy versus in-plane momentum for HE perovskite (Bottom). (d) Integrated PL signal for all k-vectors (Top); PL spectra plotted as energy versus in-plane momentum for DMPA (2) perovskite (Bottom). Both spectra are collected from perovskite microwires grown on glass substrate with PMMA grating on top for the TE polarization.

The PL signal collected from the PMMA grating for HE and DMPA (2) single crystals was investigated by Fourier space imaging and spectroscopy, by non-resonantly exciting the crystals and collecting the signal away from the excitation spot. This allows to capture only the signal coming from the propagating modes, avoiding the excitonic emission which remains just as a residue. Figure 4c and 4d show angle and energy-resolved propagating PL emission polarized along the transversal electric (TE) direction for both HE (Figure 4c) and DMPA (2) (Figure 4d) PVKs. The guided mode that is localized in plane by total internal reflection, propagates to the sharp edge of the crystal without scattering losses thanks to the high quality of the PVK slab and couples with the perovskite exciton, giving rise to a new quasi-particle called exciton-polariton. In fact, in both cases, the energy dispersions show the emergence of clear propagating modes bending at the exciton energy, which is a peculiar characteristic of the strong light-matter coupling effect.[56] Furthermore, by considering the sum of the PL emission for all k-vectors (top panels in Figure 4c and 4d), the energy of the propagating modes perfectly corresponds to the energy of the $E_2$ peak observed in PL spectrum collected at the edge of the crystals shown in Figure 3.

Our analysis demonstrates that such extra peak, only visible at the edges of the crystal, is related to the presence of waveguide optical modes hybridized with the bare exciton transition of the perovskite. The formation of waveguide exciton-polaritons in bare perovskite crystals has so far been observed for classical RP perovskites.[9] Similarly, in the case of our DJ PVK the high thickness of the crystals could lead to the formation of multiple waveguide modes.

In order to have an estimation of the light-matter coupling strength, we fitted the polariton dispersions using a two coupled oscillator model (Figure S17), extracting a Rabi splitting of 100 meV for HE and 125 meV for DMPA (2), demonstrating the high oscillator strength of these materials at RT, in good agreement with the literature.[57,58]

**Conclusions**

In this work we synthetized new 2D DJ perovskites by using organic cations with different characteristics, exploiting new design possibilities and correlating the optical properties of the developed materials to their structural parameters. The high stability and versatility of this class of semiconductors open the possibility to explore at room temperature new photophysics related to the engineering of the in plane octahedral distortion.

Furthermore, we have provided a smart and fast way to synthesize size-controlled and high-quality DJ PVK single crystal microwires, suitable to be directly integrated into nanophotonic devices. The developed microfluidic-assisted growth technique allows to obtain perovskite single crystals with high aspect ratios directly at the interface of the functional nanophotonic device. This therefore ensures ease of integration of the device components and allows for in-depth understanding of the photophysics of DJ PVK crystals.

More importantly, for the first time, single crystals of DJ PVK have been used as an active material to observe a strong light-matter interaction in a polariton waveguide.

**Data availability**

Crystallographic data of HE, DMPA (1) and DMPA (2) have been deposited at the Cambridge Crystallographic Data Centre (CCDC) with deposit number CCDC 2253676, CCDC 2253677 and CCDC 2253678, respectively, and can be obtained free of charge *via* www.ccdc.cam.ac.uk/structures.

**Acknowledgments**

The authors gratefully thank P. Cazzato, S. Carallo, F. Baldassarre and R. Lassandro for their valuable technical assistance support and Jonathan Wright for the provision of in-house time at ID11. This work was supported by the Italian Ministry of University (MIUR) for funding through the PRIN project “Interacting Photons in Polariton Circuits” — INPhoPOL (grant 2017P9FJBS), the project “Hardware implementation of a polariton neural network for neuromorphic computing” – Joint Bilateral Agreement CNR-RFBR (Russian Foundation for Basic Research) – Triennal Program 2021–2023, the MIUR project “ECOTEC - ECO-sustainable and intelligent fibers and fabrics for TEChnic clothing”, PON « R&I» 2014–2020, project N° ARS01_00951, CUP B66C18000300005, the MAECI project “Novel photonic platform for neuromorphic computing”, Joint Bilateral Project Italia-Polonia, 2022-2023 and the project “TECNOMED - Tecnopolo di Nanotecnologia e Fotonica per la Medicina di Precisione”, (Ministry of University and Scientific Research (MIUR) Decreto Direttoriale n. 3449 del 4/12/2017, CUP B83B17000010001), by the European Union - NextGeneration EU, "Integrated infrastructure initiative in Photonic and Quantum Sciences" - I-PHOQS [IR0000016, ID D2B8D520, CUP B53C22001750006], by the National and Quantum Science Technology (NQSTI) – Spoke 2 (CUP B53C22004180005, PE0000023), by the European Union - NextGeneration EU project “Network 4 Energy Sustainable Transition – NEST” (Project code PE0000021, CUP B53C22004060006, Concession Decree No. 1561 of 11.10.2022 adopted by Ministero dell’Università e della Ricerca), and by the ERC Consolidator project no. 101045746 — HYNANOSTORE. This work was (partly) supported by the Joint Bilateral Agreement CNR/FAPESP (Brazil) 2022-2023 (CUP B57G22000240001).

**Conflict of interest**

The authors declare no conflict of interest.

# Supporting Information

## Experimental details

*Chemicals and reagents*

Hydroiodic acid (HI), Hexamethylenediamine (HE), 3-(Dimethylamino)-1-propylamine) (DMPA), Hydrophosphorous acid ($H_3PO_2$), Diethyl Ether (DE), Dimethyl sulfoxide (DMSO), *N*,*N*-Dimethylformamide (DMF), Acetone, Isopropanol (IPA), Dichloromethane (DCM), Lead Oxide (PbO) and Polydimethylsiloxane (PDMS) Sylgard 184 were purchased from Dow-Corning. 4-fluoro-phenethylammonium iodide (PEAI-F) was purchased from GreatCell Solar. Lead (II) iodide ($PbI_2$) was purchased from Alfa Aesar. γ-butyrolactone (GBL) was purchased from TCI. All chemicals were used as received without any further purification.

*$HEPbI_4$ and $DMPAPbI_4$ synthesis*

Figure S1 shows a sketch of the solution-based growth method synthesis[1,2] used for HE and DMPA perovskites. Lead oxide (223.19 mg, 1 mmol) was mixed with hydroiodic acid 57% (3 ml) and 0.6 ml of 50% aqueous $H_3PO_2$ by heating under stirring for 10 minutes at 90°C. When a clear yellow solution was obtained, an amount of 290.53 mg of HE (2.5 mmol) or 255.45 mg of DMPA (2.5 mmol) was added. By slow cooling at room temperature, we obtained high quality single crystals of the desired perovskite. Then, the resulting precipitate was washed with diethyl ether. The procedure was repeated several times until all the HI was removed. The micrometer-size single crystals obtained were finally dried in vacuum for 12h on filter paper and then transferred onto the final substrate. High quality and millimeter – sized perovskite single crystals were obtained (Figure S2).

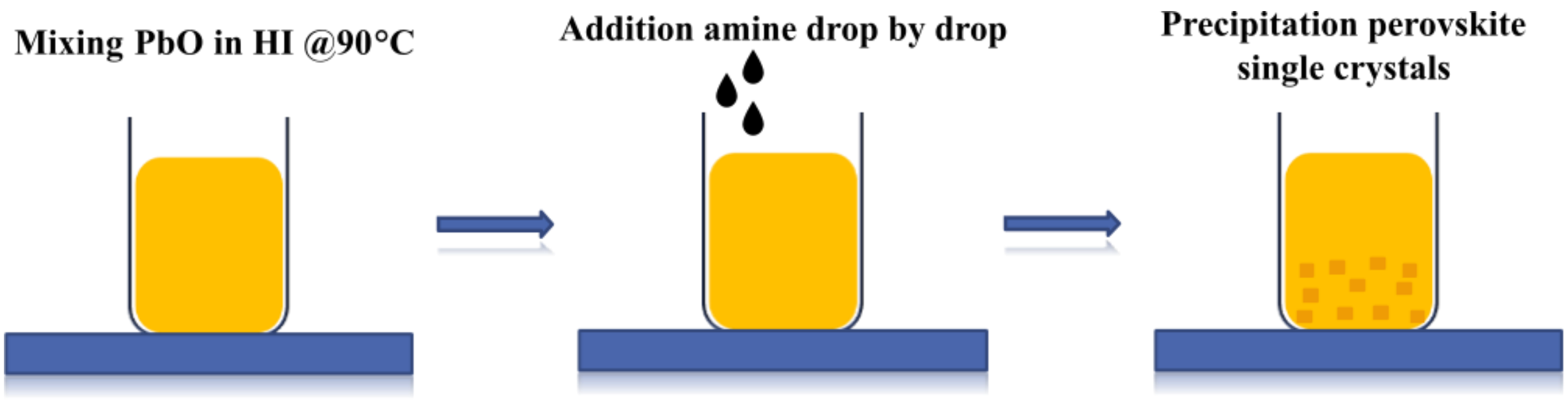

**Figure S1**. Sketch of the DJ perovskites synthetic process.

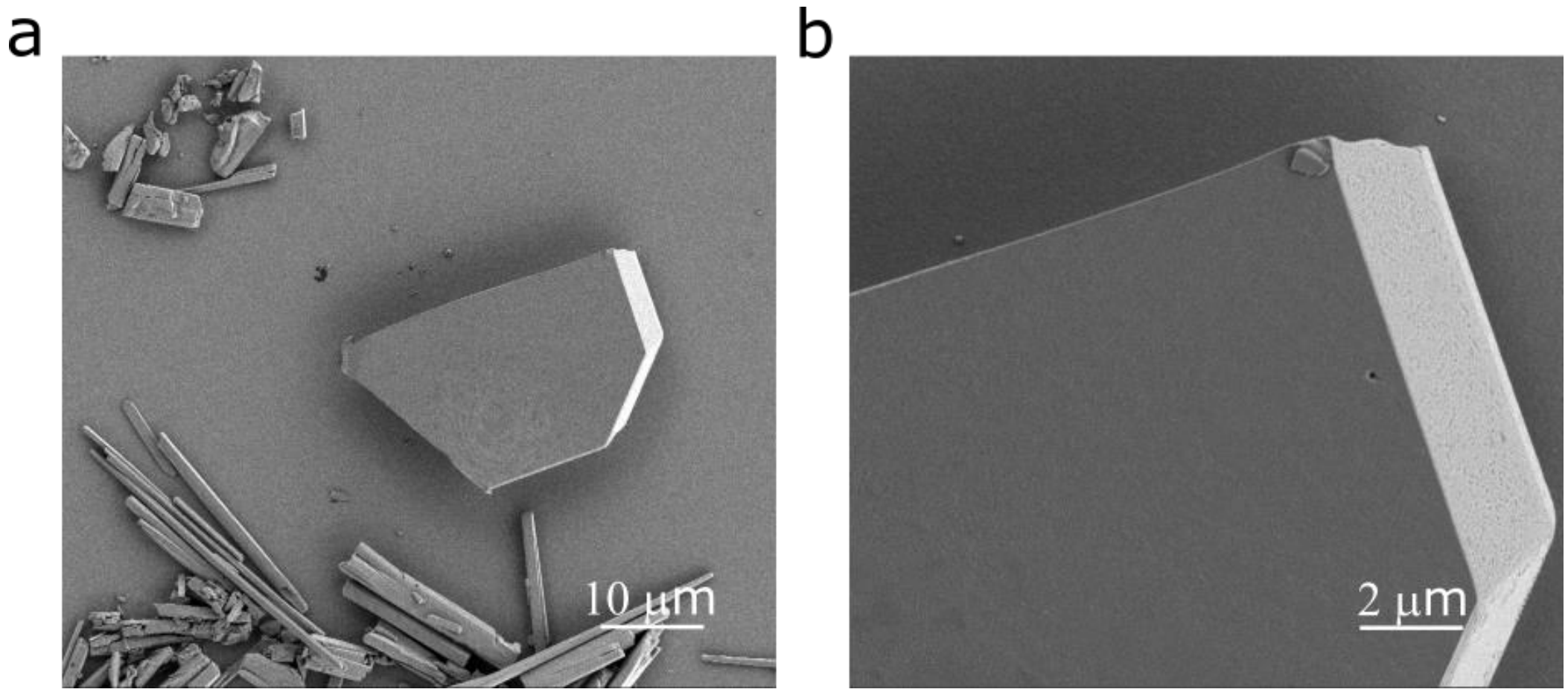

**Figure S2.** (a-b) SEM images of HE single crystals obtained through precipitation.

*(4-F-PEAI)$_2$PbI$_4$ synthesis*

Thin and high quality PEAI-F single crystals on glass substrate were obtained by using an Antisolvent vapor assisted crystallization.[3,4] Precursor solution was prepared in a $N_2$—filled glovebox: 4-fluoro-phenethylammonium iodide (267.08 mg, 1mmol) and lead iodide (230.5 mg, 0.5 mmol) were dissolved in 1 mL of γ-butyrolactone, in order to obtain a 0.5 M solution that was stirred at 70 °C for 1 h, until a clear yellow solution was obtained. Then, 3 µL of the perovskite solution was deposited on a glass substrate and then covered by another glass substrate. A small vial containing the antisolvent (Dichloromethane, 2 mL) was placed on the top of the substrates, which are closed then in a Teflon vial for 12 hours at room temperature.

**XRD measurements**

*Synchrotron micro-diffraction*

Single-crystal X-ray diffraction data for HE and DMPA (2) were collected at the beamline 1D11 at the European Synchrotron Radiation Facility (ESRF) in Grenoble, France**[5]** using a monochromatic beam produced by a bent Si (111) Laue–Laue double-crystal monochromator (≈42.26 keV, wavelength λ= 0.29339 Å for HE and ≈43.47 keV, wavelength λ = 0.2852 Å for DMPA (2), relative bandwidth $\Delta\lambda / \lambda \approx 10^{-3}$). The beam was focused to 1 µm x 1 µm in the horizontal and vertical direction with a Si refractive compound lenses system**[6]**. Main experimental details are summarized in Table S1. The crystals were centred in the beam with the help of the standard deviation signal recorded by the diffraction camera. The diffracted signals have been collected at ambient temperature (T = 293 K) by a photon counting detector (Eiger2 4M, 75 µm pixel size) placed at 119.06 mm from the sample. The sample-to-detector-distance and detector tilt have been calibrated using a $CeO_2$ powder. A series of images (typically 360) have been collected using a continuous rotation scan with an exposure time of 0.5 s. Images were then converted into the "Esperanto" format using the Eiger2esperanto script, which is part of FabIO[7] version 0.11 in order to be processed with the Crysalis software.[8] Bragg peaks were indexed, and their intensities were integrated and corrected for Lorentz polarization effects, using the Crysalis package. The semi-empirical ABSPACK routine implemented in Crysalis applied scaling and correction for absorption.

In the case of DMPA (1), single-crystal X-ray diffraction data measurements were carried out at the beamline PXII (X10SA-PXII, https://www.psi.ch/en/sls/pxii) at the Swiss Light Source (SLS), Villigen, Switzerland, using an EIGER2 16M detector. Data collection was performed at ambient temperature (T = 293 K) on a selected crystal mounted on litholoops (Molecular Dimensions), in steps of 0.2 ° at a speed of 10 °/s with the EIGER2 16M detector operated in a continuous/shutterless data collection mode. Complete data were obtained by collecting 360° ω scan at a beam energy of ≈19.07 keV (λ = 0.64997Å), focus size 70 × 20 $\mu m^2$. The principal experimental details are given in Table S1. Diffraction data were processed by XDS,[9] a software organized in eight subroutines able to perform the main data reduction steps; the integrated intensities were scaled and corrected for absorption effects by the XSCALE subroutine.

Crystal structure was solved by Direct Methods[10] using *SIR2019*[11] and refined by *SHELXL2014/7*[12]. Additional applied computer programs were: *Mercury*[13] and *VESTA*[14] for molecular graphics; *WinGX*[15] and *publCIF*[16] for preparing the published material.

*Structure solution and refinement*

The structure solution process for the HE, DMPA (2) and DMPA (1) crystals was carried out by *SIR2019*,[11] a package able to determine the space group and the crystal structure by exploiting minimal prior information concerning the unit cell parameters, the observed diffraction intensities and the expected chemical formula. The space group determination step was performed by considering the Laue group compatible with the monoclinic (HE and DMPA (2)) or the orthorhombic (DMPA (1)) unit cell and by calculating, for each possible extinction symbol, a suitable probability value *via* a statistical analysis based on the experimental intensities. The structure model was completed by locating the missing non-H light atoms *via* difference Fourier synthesis and refined using full-matrix least-squares techniques by *SHELXL2014/7*.[12] All non-hydrogen atoms were refined anisotropically. The H atoms were placed at calculated *positions* and their atomic coordinates were refined according to a riding model, except for one of the H atoms of DMPA (2), *i.e.*, HN1, bonded to the N1 nitrogen atom;

HN1 was located *via* a careful inspection of the electron density map calculated by difference Fourier synthesis and its fractional coordinates were freely refined. The constraints on the isotropic $U$ value of H atoms bonded to C and N atoms, were $U_{iso}$(H)=1.2 $U_{eq}$(C) and $U_{iso}$(H)=1.5 $U_{eq}$(N), respectively.

Main details on data collection and structure refinement are given in Table S1.

| *Crystal* | *HE* | *DMPA (2)* | *DMPA (1)* |
|---|---|---|---|
| Chemical formula | $C_6H_{18}N_2PbI_4$ | $C_5H_{16}N_2PbI_4$ | $C_5H_{16}N_2PbI_4$ |
| *Data collection* | | | |
| Diffractometer | ID11 nanoscope | ID11 nanoscope | X10SA-PXII beamline, single-axis |
| Absorption correction | Empirical (using intensity measurements) SCALE3 ABSPACK; Rigaku Oxford Diffraction, (2015) | Empirical (using intensity measurements) SCALE3 ABSPACK; Rigaku Oxford Diffraction, (2015) | Empirical (using intensity measurements) |
| $T_{min}$, $T_{max}$ | 0.34, 1.00 | 0.32, 1.0 | 0.33, 0.99 |
| No. of measured, independent and observed [$I > 2\sigma(I)$] reflections | 14269, 2126, 1867 | 18428, 3336, 2810 | 27262, 4401, 4149 |
| $R_{int}$ | 0.044 | 0.070 | 0.050 |
| $(\sin \theta/\lambda)_{max}$ (Å$^{-1}$) | 0.667 | 0.625 | 0.685 |
| *Refinement* | | | |
| $R[F^2 > 2\sigma(F^2)]$, $wR(F^2)$, $S$ | 0.042, 0.114, 1.10 | 0.037, 0.094, 1.04 | 0.027, 0.073, 1.06 |
| No. of reflections | 2126 | 3336 | 4401 |
| No. of parameters | 62 | 115 | 112 |
| H-atom treatment | H-atom parameters constrained | H atoms treated by a mixture of independent and constrained refinement | H-atom parameters constrained |
| $\Delta\rho_{max}$, $\Delta\rho_{min}$ (e Å$^{-3}$) | 4.03, -1.21 | 2.25, −1.66 | 1.46, −1.28 |
| CCDC number | 2253676 | 2253678 | 2253677 |

**Table S1.** HE, DMPA (2), DMPA (1): data collection and refinement details.

For each compound, two additional tables (*i.e.*, Table S2 and S3 for HE, Table S4 and S5 for DMPA (2) and Table S6 and S7 for DMPA (1)) supplying refined fractional atomic coordinates, displacement parameters, bond distances and angles, torsion angles and hydrogen bonds, are provided.

The Pb-I bond lengths in the $PbI_6$ octahedron belonged to the intervals 3.1859(6) - 3.2269(5) Å for HE (see Table S2), 3.1789(5) - 3.2467(5) Å for DMPA (2) (see Table S4) and 3.1238(4) - 3.3015(4) Å for DMPA (1) (see Table S6). The ranges of the $I_{ap}$-Pb-$I_{eq}$ octahedral bond angles

were 87.665(15) ° - 92.335(15) ° for HE (see Table S2), 83.150(16) °- 101.236(17)° for DMPA (2) (see Table S4) and 83.731(11)° - 94.899(16)° for DMPA (1) (see Table S6), with $I_{ap}$ and $I_{eq}$ the apex and equatorial I atoms of the octahedron, respectively.

In the case of HE, a view of the asymmetric unit and its local environment is given in Figure S3, showing also the main directional H-bonds, *i.e.*, N—H··· I intra and intermolecular interactions, for which the hydrogen-acceptor distances vary in the range 2.76-3.26 Å (see Table S3 and Figure S3b) and their influence in the equatorial distortion (see Figure S4). This results in a high level of octahedral distortion in HE compound with a $Pb_{eq}$–I–$Pb_{eq}$ angle of 148.63° (see Figure S3b and Table S2), far from the ideal value (180.0°).

In the case of DMPA (2), the view of the asymmetric unit and its local environment is provided in Figure S5, showing the equatorial distortion $Pb_{eq}$–I–$Pb_{eq}$ angle (*i.e.*, 156.83°, see Figure S5b, Table S4). The main intermolecular interactions consist of N—H··· I and C—H··· I interactions for which the minimum and maximum values of the hydrogen-acceptor distances are 2.77 and 3.28 Å, respectively (see Table S5). The tilting of adjacent polyhedra is clearly visible in Figure S6.

In the case of DMPA (1), the view of the asymmetric unit and its local environment is given in Figure S7, showing, in addition to the equatorial distortion $Pb_{eq}$–I–$Pb_{eq}$ angle (*i.e.*, 163.13°, see Figure S7b and Table S6) also some directional H-bonds (*i.e.*, three intramolecular interactions, for which the hydrogen-acceptor distances vary in the range 2.88-3.31 Å; see Table S7 for the list of all the main H-bonds). The tilting of adjacent polyhedra is clearly shown in Figure S8.

*Computing details*

Program used for cell refinement and data reduction CrysAlis PRO (Rigaku Oxford Diffraction, 2015); program used to solve structure: *SIR2019* (Burla *et al.*, 2015); program used to refine structure *SHELXL2014/7* (Sheldrick, 2015); molecular graphics: *Mercury* (Macrae *et al*., 2020) and *VESTA* (Momma & Izumi, 2011); software used to prepare material for publication: *WinGX* (Farrugia, 2012) and *publCIF* (Westrip, 2010).

**Table S2.** HE: main experimental and crystallographic details.

| *Crystal data* | |
|---|---|
| $C_6H_{18}N_2PbI_4$ | $F$(000) = 724 |
| $M_r$ = 833.01 | $D_x$ = 3.169 Mg m$^{-3}$ |
| Monoclinic, $P2_1/c$ | Synchrotron radiation, λ = 0.29339 Å |
| $a$ = 11.8712 (4) Å | Cell parameters from 4356 reflections |
| $b$ = 8.4911 (2) Å | θ = 3.6–20.4° |
| $c$ = 9.0503 (3) Å | μ = 4.67 mm$^{-1}$ |
| β = 106.856 (3)° | $T$ = 293 K |
| $V$ = 873.07 (5) Å$^3$ | Block, pale yellow |
| $Z$ = 2 | 0.10 × 0.09 × 0.07 mm |

## *Data collection*

| | |
|---|---|
| ID11 nanoscope diffractometer | $R_{int}$ = 0.044 |
| Radiation source: synchrotron | $\theta_{max}$ = 11.3°, $\theta_{min}$ = 1.8° |
| rotation scans | $h$ = -15→15 |
| Absorption correction: empirical (using intensity measurements) SCALE3 ABSPACK; Rigaku Oxford Diffraction, (2015) | $k$ = -11→11<br>$l$ = -12→12 |
| $T_{min}$ = 0.34, $T_{max}$ = 1.0 | |
| 14269 measured reflections | 2126 standard reflections every 0.1 reflections |
| 2126 independent reflections | intensity decay: none |
| 1867 reflections with $I > 2\sigma(I)$ | |

## *Refinement*

| | |
|---|---|
| Refinement on $F^2$ | 0 restraints |
| Least-squares matrix: full | Hydrogen site location: inferred from neighbouring sites |
| $R[F^2 > 2\sigma(F^2)]$ = 0.042 | H-atom parameters constrained |
| $wR(F^2)$ = 0.114 | $w = 1/[\sigma^2(F_o^2) + (0.0718P)^2 + 1.8531P]$<br>where $P = (F_o^2 + 2F_c^2)/3$ |
| $S$ = 1.10 | $(\Delta/\sigma)_{max}$ < 0.001 |
| 2126 reflections | $\Delta\rho_{max}$ = 4.03 e Å$^{-3}$ |
| 62 parameters | $\Delta\rho_{min}$ = -1.21 e Å$^{-3}$ |

## *Special details*

*Geometry*. All estimated standard deviations (esds), except the esd in the dihedral angle between two least squares (l.s.) planes, are estimated using the full covariance matrix. The cell esds are taken into account individually in the estimation of esds in distances, angles and torsion angles; correlations between esds in cell parameters are only used when they are defined by crystal symmetry. An approximate (isotropic) treatment of cell esds is used for estimating esds involving l.s. planes.

## *Fractional atomic coordinates and isotropic or equivalent isotropic displacement parameters (Å$^2$)*

| | $x$ | $y$ | $z$ | $U_{iso}$*/$U_{eq}$ |
|---|---|---|---|---|
| Pb1 | 0.0000 | 0.0000 | 0.0000 | 0.04071 (15) |
| I1 | 0.04945 (5) | 0.19317 (7) | 0.31951 (6) | 0.05344 (18) |
| I2 | -0.27305 (5) | 0.08024 (7) | -0.07629 (7) | 0.05551 (19) |
| C1 | -0.3399 (9) | -0.0987 (12) | 0.2875 (13) | 0.061 (2) |
| H1A | -0.3117 | -0.1990 | 0.2610 | 0.074* |
| H1B | -0.3966 | -0.0580 | 0.1955 | 0.074* |
| C2 | -0.4010 (9) | -0.1253 (12) | 0.4107 (13) | 0.061 (2) |
| H2A | -0.4590 | -0.2081 | 0.3766 | 0.073* |
| H2B | -0.3433 | -0.1617 | 0.5039 | 0.073* |
| C3 | -0.4611 (8) | 0.0180 (10) | 0.4491 (11) | 0.0509 (19) |
| H3A | -0.5085 | 0.0656 | 0.3540 | 0.061* |
| H3B | -0.4018 | 0.0940 | 0.5011 | 0.061* |
| N1 | -0.2402 (8) | 0.0113 (10) | 0.3349 (12) | 0.063 (2) |
| H1NA | -0.2003 | 0.0099 | 0.2654 | 0.095* |
| H1NB | -0.1929 | -0.0175 | 0.4265 | 0.095* |
| H1NC | -0.2670 | 0.1082 | 0.3414 | 0.095* |

*Atomic displacement parameters (Å$^2$)*

| | $U^{11}$ | $U^{22}$ | $U^{33}$ | $U^{12}$ | $U^{13}$ | $U^{23}$ |
|---|---|---|---|---|---|---|
| Pb1 | 0.0516 (2) | 0.0356 (2) | 0.0403 (2) | 0.00085 (13) | 0.02174 (17) | -0.00031 (13) |
| I1 | 0.0667 (4) | 0.0468 (3) | 0.0515 (3) | -0.0017 (2) | 0.0245 (3) | -0.0169 (2) |
| I2 | 0.0527 (3) | 0.0516 (3) | 0.0640 (4) | 0.0042 (2) | 0.0199 (3) | 0.0069 (2) |
| C1 | 0.067 (6) | 0.059 (5) | 0.071 (6) | -0.008 (4) | 0.040 (5) | -0.016 (4) |
| C2 | 0.061 (5) | 0.059 (5) | 0.073 (6) | -0.001 (4) | 0.037 (5) | -0.003 (5) |
| C3 | 0.052 (4) | 0.052 (5) | 0.054 (5) | -0.007 (3) | 0.024 (4) | -0.003 (3) |
| N1 | 0.066 (5) | 0.061 (5) | 0.074 (6) | 0.005 (3) | 0.038 (5) | 0.004 (4) |

*Geometric parameters (Å, º)*

| | | | |
|---|---|---|---|
| Pb1—I2 | 3.1859 (6) | C1—H1B | 0.9700 |
| Pb1—I2$^{i}$ | 3.1860 (6) | C2—C3 | 1.501 (13) |
| Pb1—I1$^{ii}$ | 3.2181 (5) | C2—H2A | 0.9700 |
| Pb1—I1$^{iii}$ | 3.2181 (5) | C2—H2B | 0.9700 |
| Pb1—I1 | 3.2268 (5) | C3—C3$^{v}$ | 1.513 (17) |
| Pb1—I1$^{i}$ | 3.2269 (5) | C3—H3A | 0.9700 |
| I1—Pb1$^{iv}$ | 3.2182 (5) | C3—H3B | 0.9700 |
| C1—N1 | 1.471 (13) | N1—H1NA | 0.8900 |
| C1—C2 | 1.513 (13) | N1—H1NB | 0.8900 |
| C1—H1A | 0.9700 | N1—H1NC | 0.8900 |
| | | | |
| I2—Pb1—I2$^{i}$ | 180.0 | C2—C1—H1B | 108.9 |
| I2—Pb1—I1$^{ii}$ | 92.335 (15) | H1A—C1—H1B | 107.7 |
| I2$^{i}$—Pb1—I1$^{ii}$ | 87.665 (15) | C3—C2—C1 | 114.3 (9) |
| I2—Pb1—I1$^{iii}$ | 87.665 (15) | C3—C2—H2A | 108.7 |
| I2$^{i}$—Pb1—I1$^{iii}$ | 92.335 (15) | C1—C2—H2A | 108.7 |
| I1$^{ii}$—Pb1—I1$^{iii}$ | 180.000 (12) | C3—C2—H2B | 108.7 |
| I2—Pb1—I1 | 89.701 (16) | C1—C2—H2B | 108.7 |
| I2$^{i}$—Pb1—I1 | 90.300 (16) | H2A—C2—H2B | 107.6 |
| I1$^{ii}$—Pb1—I1 | 91.764 (6) | C2—C3—C3$^{v}$ | 113.1 (9) |
| I1$^{iii}$—Pb1—I1 | 88.236 (6) | C2—C3—H3A | 109.0 |
| I2—Pb1—I1$^{i}$ | 90.299 (16) | C3$^{v}$—C3—H3A | 109.0 |
| I2$^{i}$—Pb1—I1$^{i}$ | 89.700 (16) | C2—C3—H3B | 109.0 |
| I1$^{ii}$—Pb1—I1$^{i}$ | 88.236 (5) | C3$^{v}$—C3—H3B | 109.0 |
| I1$^{iii}$—Pb1—I1$^{i}$ | 91.764 (5) | H3A—C3—H3B | 107.8 |
| I1—Pb1—I1$^{i}$ | 180.0 | C1—N1—H1NA | 109.5 |
| $Pb1^{iv}$—I1—Pb1 | 148.63 (2) | C1—N1—H1NB | 109.5 |
| N1—C1—C2 | 113.4 (8) | H1NA—N1—H1NB | 109.5 |
| N1—C1—H1A | 108.9 | C1—N1—H1NC | 109.5 |
| C2—C1—H1A | 108.9 | H1NA—N1—H1NC | 109.5 |
| N1—C1—H1B | 108.9 | H1NB—N1—H1NC | 109.5 |
| | | | |
| N1—C1—C2—C3 | 65.4 (12) | C1—C2—C3—$C3^{v}$ | 169.8 (10) |

Symmetry codes: (i) -*x*, -*y*, -*z*; (ii) *x*, -*y*+1/2, *z*-1/2; (iii) -*x*, *y*-1/2, -*z*+1/2; (iv) -*x*, *y*+1/2, -*z*+1/2; (v) -*x*-1, -*y*, -*z*+1.

**Table S3.** HE: Hydrogen-bond geometry (Å,°)

| *D*—H⋯*A* | *D*—H | H⋯*A* | *D*⋯*A* | *D*—H⋯*A* |
|---|---|---|---|---|
| N1—H1NA⋯I1 | 0.89 | 3.26 | 3.808 (9) | 122 |
| N1—H1NA⋯I2 | 0.89 | 3.02 | 3.674 (10) | 132 |
| N1—H1NB⋯$I1^{vi}$ | 0.89 | 2.85 | 3.717 (10) | 164 |
| N1—H1NC⋯$I2^{vii}$ | 0.89 | 2.76 | 3.607 (8) | 161 |

Symmetry codes: (vi) -*x*, -*y*, -*z*+1; (vii) *x*, -*y*+1/2, *z*+1/2.

**Table S4.** DMPA (2): main experimental and crystallographic details.

| *Crystal data* | |
|---|---|
| $C_5H_{16}N_2PbI_4$ | $F(000) = 1416$ |
| $M_r = 818.99$ | $D_x = 3.268$ Mg $m^{-3}$ |
| Monoclinic, $P2_1/c$ | Synchrotron radiation, $\lambda = 0.2852$ Å |
| $a = 11.2176$ (4) Å | Cell parameters from 4122 reflections |
| $b = 12.5073$ (3) Å | $\theta = 4.0$–21.0° |
| $c = 12.8819$ (3) Å | $\mu = 4.55$ $mm^{-1}$ |
| $\beta = 112.929$ (3)° | $T = 293$ K |
| $V = 1664.55$ (9) $Å^3$ | Prism, red |
| $Z = 4$ | 0.10 × 0.10 × 0.03 mm |

| *Data collection* | |
|---|---|
| ID11 nanoscope diffractometer | 2810 reflections with $I > 2\sigma(I)$ |
| Radiation source: synchrotron | $R_{int} = 0.070$ |
| rotation scans | $\theta_{max} = 10.3°$, $\theta_{min} = 1.5°$ |
| Absorption correction: empirical (using intensity measurements) SCALE3 ABSPACK; Rigaku Oxford Diffraction, (2015) | $h = -14 \rightarrow 14$ |
| $T_{min} = 0.32$, $T_{max} = 1.0$ | $k = -15 \rightarrow 15$ |
| 18428 measured reflections | $l = -16 \rightarrow 16$ |

3336 independent reflections

## *Refinement*

| | |
|---|---|
| Refinement on $F^2$ | 0 restraints |
| Least-squares matrix: full | Hydrogen site location: mixed |
| $R[F^2 > 2\sigma(F^2)] = 0.037$ | H atoms treated by a mixture of independent and constrained refinement |
| $wR(F^2) = 0.094$ | $w = 1/[\sigma^2(F_o^2) + (0.0478P)^2 + 4.5716P]$ where $P = (F_o^2 + 2F_c^2)/3$ |
| $S = 1.04$ | $(\Delta/\sigma)_{max} < 0.001$ |
| 3336 reflections | $\Delta\rho_{max} = 2.25$ e Å$^{-3}$ |
| 115 parameters | $\Delta\rho_{min} = -1.66$ e Å$^{-3}$ |

## *Special details*

*Geometry*. All esds (except the esd in the dihedral angle between two l.s. planes) are estimated using the full covariance matrix. The cell esds are taken into account individually in the estimation of esds in distances, angles and torsion angles; correlations between esds in cell parameters are only used when they are defined by crystal symmetry. An approximate (isotropic) treatment of cell esds is used for estimating esds involving l.s. planes.

## *Fractional atomic coordinates and isotropic or equivalent isotropic displacement parameters (Å$^2$)*

| | $x$ | $y$ | $z$ | $U_{iso}$*/$U_{eq}$ |
|---|---|---|---|---|
| Pb1 | 0.49944 (3) | 0.25715 (2) | 0.23207 (2) | 0.03569 (11) |
| I1 | 0.51730 (6) | 0.24316 (5) | -0.01024 (4) | 0.04902 (17) |
| I2 | 0.55931 (7) | 0.00786 (4) | 0.25501 (5) | 0.05060 (18) |
| I3 | 0.80521 (6) | 0.30314 (5) | 0.33832 (5) | 0.05465 (18) |
| I4 | 0.19420 (6) | 0.22222 (6) | 0.10462 (6) | 0.05936 (19) |
| N1 | 0.1456 (8) | 0.5211 (6) | -0.1477 (7) | 0.0492 (18) |
| HN1 | 0.161 (11) | 0.565 (9) | -0.179 (9) | 0.059* |
| C3 | 0.0373 (11) | 0.4522 (9) | -0.2150 (9) | 0.063 (3) |
| H3A | 0.0141 | 0.4070 | -0.1646 | 0.076* |
| H3B | 0.0652 | 0.4060 | -0.2618 | 0.076* |
| C4 | 0.2681 (11) | 0.4617 (9) | -0.0997 (9) | 0.064 (3) |
| H4A | 0.2953 | 0.4415 | -0.1590 | 0.096* |
| H4B | 0.2559 | 0.3987 | -0.0625 | 0.096* |
| H4C | 0.3330 | 0.5061 | -0.0463 | 0.096* |
| C5 | -0.0845 (12) | 0.5155 (9) | -0.2917 (10) | 0.070 (3) |
| H5A | -0.1291 | 0.5435 | -0.2465 | 0.084* |
| H5B | -0.0581 | 0.5755 | -0.3256 | 0.084* |
| C6 | -0.1711 (12) | 0.4478 (10) | -0.3792 (9) | 0.077 (3) |
| H6A | -0.2009 | 0.3904 | -0.3447 | 0.093* |
| H6B | -0.1238 | 0.4158 | -0.4203 | 0.093* |
| C7 | 0.1265 (12) | 0.5831 (11) | -0.0592 (11) | 0.086 (4) |
| H7A | 0.1966 | 0.6327 | -0.0272 | 0.129* |
| H7B | 0.1240 | 0.5356 | -0.0015 | 0.129* |
| H7C | 0.0463 | 0.6216 | -0.0908 | 0.129* |
| N2 | -0.2848 (10) | 0.5053 (6) | -0.4597 (8) | 0.067 (2) |
| HN2B | -0.2589 | 0.5521 | -0.4987 | 0.101* |
| HN2C | -0.3246 | 0.5399 | -0.4220 | 0.101* |
| HN2A | -0.3392 | 0.4587 | -0.5067 | 0.101* |

*Atomic displacement parameters (Å$^2$)*

| | $U^{11}$ | $U^{22}$ | $U^{33}$ | $U^{12}$ | $U^{13}$ | $U^{23}$ |
|---|---|---|---|---|---|---|
| Pb1 | 0.04126 (19) | 0.03428 (17) | 0.03304 (18) | -0.00026 (11) | 0.01613 (13) | -0.00059 (10) |
| I1 | 0.0626 (4) | 0.0582 (3) | 0.0322 (3) | 0.0094 (3) | 0.0249 (3) | 0.0032 (2) |
| I2 | 0.0647 (4) | 0.0297 (3) | 0.0552 (3) | -0.0009 (2) | 0.0210 (3) | 0.0001 (2) |
| I3 | 0.0428 (3) | 0.0681 (4) | 0.0493 (3) | -0.0075 (3) | 0.0138 (3) | 0.0091 (3) |
| I4 | 0.0395 (3) | 0.0673 (4) | 0.0699 (4) | 0.0004 (3) | 0.0198 (3) | 0.0019 (3) |
| N1 | 0.049 (5) | 0.050 (4) | 0.051 (4) | 0.001 (4) | 0.022 (4) | 0.000 (3) |
| C3 | 0.060 (7) | 0.065 (6) | 0.065 (6) | 0.000 (5) | 0.026 (5) | -0.012 (5) |
| C4 | 0.059 (7) | 0.065 (6) | 0.063 (6) | 0.010 (5) | 0.017 (5) | -0.004 (5) |
| C5 | 0.053 (7) | 0.076 (7) | 0.080 (7) | 0.006 (5) | 0.026 (6) | -0.009 (6) |
| C6 | 0.081 (9) | 0.081 (7) | 0.052 (6) | 0.003 (6) | 0.007 (5) | -0.010 (5) |
| C7 | 0.060 (7) | 0.110 (10) | 0.088 (8) | 0.006 (7) | 0.030 (6) | -0.047 (7) |
| N2 | 0.063 (6) | 0.070 (6) | 0.066 (5) | -0.009 (4) | 0.021 (5) | 0.005 (4) |

*Geometric parameters (Å, º)*

| | | | |
|---|---|---|---|
| Pb1—I2 | 3.1789 (5) | C4—H4A | 0.9600 |
| Pb1—I4 | 3.1985 (7) | C4—H4B | 0.9600 |
| Pb1—I1 | 3.2096 (5) | C4—H4C | 0.9600 |
| Pb1—I3 | 3.2110 (7) | C5—C6 | 1.442 (16) |
| Pb1—I2$^{i}$ | 3.2218 (5) | C5—H5A | 0.9700 |
| Pb1—I1$^{ii}$ | 3.2467 (5) | C5—H5B | 0.9700 |
| I1—Pb1$^{iii}$ | 3.2467 (5) | C6—N2 | 1.480 (15) |
| I2—Pb1$^{iv}$ | 3.2218 (5) | C6—H6A | 0.9700 |
| N1—C7 | 1.463 (13) | C6—H6B | 0.9700 |
| N1—C3 | 1.466 (13) | C7—H7A | 0.9600 |
| N1—C4 | 1.469 (13) | C7—H7B | 0.9600 |
| N1—HN1 | 0.74 (10) | C7—H7C | 0.9600 |
| C3—C5 | 1.556 (16) | N2—HN2B | 0.8900 |
| C3—H3A | 0.9700 | N2—HN2C | 0.8900 |
| C3—H3B | 0.9700 | N2—HN2A | 0.8900 |
| | | | |
| I2—Pb1—I4 | 93.374 (19) | N1—C4—H4B | 109.5 |
| I2—Pb1—I1 | 86.901 (16) | H4A—C4—H4B | 109.5 |
| I4—Pb1—I1 | 87.323 (19) | N1—C4—H4C | 109.5 |
| I2—Pb1—I3 | 89.093 (19) | H4A—C4—H4C | 109.5 |
| I4—Pb1—I3 | 174.492 (19) | H4B—C4—H4C | 109.5 |
| I1—Pb1—I3 | 87.895 (18) | C6—C5—C3 | 110.8 (10) |
| I2—Pb1—I2$^{i}$ | 171.838 (10) | C6—C5—H5A | 109.5 |
| I4—Pb1—I2$^{i}$ | 87.699 (19) | C3—C5—H5A | 109.5 |
| I1—Pb1—I2$^{i}$ | 101.236 (17) | C6—C5—H5B | 109.5 |
| I3—Pb1—I2$^{i}$ | 90.555 (19) | C3—C5—H5B | 109.5 |
| I2—Pb1—I1$^{ii}$ | 88.690 (16) | H5A—C5—H5B | 108.1 |
| I4—Pb1—I1$^{ii}$ | 98.740 (19) | C5—C6—N2 | 113.3 (10) |
| I1—Pb1—I1$^{ii}$ | 172.71 (2) | C5—C6—H6A | 108.9 |
| I3—Pb1—I1$^{ii}$ | 86.229 (18) | N2—C6—H6A | 108.9 |
| I2$^{i}$—Pb1—I1$^{ii}$ | 83.150 (16) | C5—C6—H6B | 108.9 |
| Pb1—I1—Pb1$^{iii}$ | 172.77 (2) | N2—C6—H6B | 108.9 |
| Pb1—I2—Pb1$^{iv}$ | 156.83 (3) | H6A—C6—H6B | 107.7 |
| C7—N1—C3 | 115.7 (9) | N1—C7—H7A | 109.5 |
| C7—N1—C4 | 109.7 (8) | N1—C7—H7B | 109.5 |
| C3—N1—C4 | 112.0 (8) | H7A—C7—H7B | 109.5 |
| C7—N1—HN1 | 100 (9) | N1—C7—H7C | 109.5 |
| C3—N1—HN1 | 115 (9) | H7A—C7—H7C | 109.5 |
| C4—N1—HN1 | 103 (9) | H7B—C7—H7C | 109.5 |
| N1—C3—C5 | 113.4 (9) | C6—N2—HN2B | 109.5 |
| N1—C3—H3A | 108.9 | C6—N2—HN2C | 109.5 |
| C5—C3—H3A | 108.9 | HN2B—N2—HN2C | 109.5 |
| N1—C3—H3B | 108.9 | C6—N2—HN2A | 109.5 |
| C5—C3—H3B | 108.9 | HN2B—N2—HN2A | 109.5 |
| H3A—C3—H3B | 107.7 | HN2C—N2—HN2A | 109.5 |
| N1—C4—H4A | 109.5 | | |
| | | | |
| C7—N1—C3—C5 | 68.3 (13) | N1—C3—C5—C6 | 162.3 (10) |
| C4—N1—C3—C5 | -165.0 (10) | C3—C5—C6—N2 | -176.1 (10) |

Symmetry codes: (i) -*x*+1, *y*+1/2, -*z*+1/2; (ii) *x*, -*y*+1/2, *z*+1/2; (iii) *x*, -*y*+1/2, *z*-1/2; (iv) -*x*+1, *y*-1/2, -*z*+1/2.

**Table S5.** DMPA (2): Hydrogen-bond geometry (Å, °)

| *D*—H···*A* | *D*—H | H···*A* | *D*···*A* | *D*—H···*A* |
|---|---|---|---|---|
| N1—H*N*1···I3$^{i}$ | 0.74 (10) | 2.77 (11) | 3.496 (8) | 167 (11) |
| C3—H3*B*···I4$^{ii}$ | 0.97 | 3.09 | 4.055 (11) | 171 |
| C5—H5*B*···I3$^{i}$ | 0.97 | 3.28 | 4.097 (13) | 143 |
| C6—H6*B*···I3$^{iii}$ | 0.97 | 3.22 | 3.979 (11) | 137 |
| C7—H7*A*···I3$^{iv}$ | 0.96 | 3.24 | 3.815 (11) | 120 |
| C7—H7*C*···I4$^{v}$ | 0.96 | 3.28 | 4.192 (12) | 160 |
| N2—H*N*2*B*···I4$^{vi}$ | 0.89 | 2.77 | 3.654 (8) | 171 |
| N2—H*N*2*C*···I1$^{vi}$ | 0.89 | 3.23 | 3.875 (9) | 131 |
| N2—H*N*2*C*···I2$^{vii}$ | 0.89 | 3.09 | 3.808 (10) | 139 |
| N2—H*N*2*A*···I1$^{vii}$ | 0.89 | 2.99 | 3.727 (9) | 142 |
| N2—H*N*2*A*···I2$^{vi}$ | 0.89 | 3.25 | 3.851 (9) | 127 |

Symmetry codes: (i) -*x*+1, -*y*+1, -*z*; (ii) *x*, -*y*+1/2, *z*-1/2; (iii) *x*-1, *y*, *z*-1; (iv) -*x*+1, *y*+1/2, -*z*+1/2; (v) -*x*, -*y*+1, -*z*; (vi) -*x*, *y*+1/2, -*z*-1/2; (vii) *x*-1, -*y*+1/2, *z*-1/2.

**Table S6.** DMPA (1): main experimental and crystallographic details.

| *Crystal data* | |
|---|---|
| $C_5H_{16}I_4N_2Pb$ | $D_x$ = 3.297 Mg m$^{-3}$ |
| $M_r$ = 818.99 | Synchrotron radiation, λ = 0.64997 Å |
| Orthorhombic, *Pbca* | Cell parameters from 15256 reflections |
| *a* = 18.406 (1) Å | θ = 0.8–24.9° |
| *b* = 8.650 (2) Å | μ = 13.88 mm$^{-1}$ |
| *c* = 20.726 (1) Å | *T* = 293 K |
| *V* = 3299.8 (8) Å$^3$ | Thin plate, red |
| *Z* = 8 | 0.03 × 0.02 × 0.01 mm |
| *F*(000) = 2832 | |
| *Data collection* | |
| X10SA-PXII beamline, single-axis diffractometer | 4149 reflections with $I > 2\sigma(I)$ |
| Radiation source: synchrotron | $R_{int}$ = 0.050 |
| ω scan, shutter–less continuous rotation | $\theta_{max}$ = 26.4°, $\theta_{min}$ = 2.0° |
| Absorption correction: empirical (using intensity measurements) | *h* = -25→24 |
| $T_{min}$ = 0.33, $T_{max}$ = 0.99 | *k* = -11→11 |
| 27262 measured reflections | *l* = -28→28 |
| 4401 independent reflections | |
| *Refinement* | |
| Refinement on $F^2$ | 0 restraints |
| Least-squares matrix: full | Hydrogen site location: inferred from neighbouring sites |
| $R[F^2 > 2\sigma(F^2)]$ = 0.027 | H-atom parameters constrained |
| $wR(F^2)$ = 0.073 | $w = 1/[\sigma^2(F_o^2) + (0.0388P)^2 + 4.574P]$ where $P = (F_o^2 + 2F_c^2)/3$ |
| *S* = 1.06 | $(\Delta/\sigma)_{max}$ = 0.002 |
| 4401 reflections | $\Delta\rho_{max}$ = 1.46 e Å$^{-3}$ |
| 112 parameters | $\Delta\rho_{min}$ = -1.28 e Å$^{-3}$ |

## *Special details*

*Geometry*. All esds (except the esd in the dihedral angle between two l.s. planes) are estimated using the full covariance matrix. The cell esds are taken into account individually in the estimation of esds in distances, angles and torsion angles; correlations between esds in cell parameters are only used when they are defined by crystal symmetry. An approximate (isotropic) treatment of cell esds is used for estimating esds involving l.s. planes.

## *Fractional atomic coordinates and isotropic or equivalent isotropic displacement parameters (Å$^2$)*

| | *x* | *y* | *z* | $U_{iso}$*/$U_{eq}$ |
|---|---|---|---|---|
| Pb1 | 0.37723 (2) | 0.37067 (2) | 0.23012 (2) | 0.04002 (6) |
| I1 | 0.39912 (2) | 0.37664 (4) | 0.08068 (2) | 0.05919 (10) |
| I2 | 0.51056 (2) | 0.59165 (4) | 0.25600 (2) | 0.05961 (10) |
| I3 | 0.34568 (2) | 0.33551 (5) | 0.38625 (2) | 0.06824 (11) |
| I4 | 0.26526 (2) | 0.64975 (4) | 0.21626 (2) | 0.06071 (10) |
| N1 | 0.3772 (2) | 0.7922 (6) | 0.0523 (2) | 0.0616 (11) |
| HN1 | 0.3703 | 0.7007 | 0.0606 | 0.074* |
| C1 | 0.4090 (3) | 0.7797 (9) | -0.0128 (3) | 0.0773 (17) |
| H1A | 0.4565 | 0.7315 | -0.0102 | 0.093* |
| H1B | 0.4150 | 0.8820 | -0.0312 | 0.093* |
| C2 | 0.3584 (4) | 0.6812 (11) | -0.0567 (3) | 0.088 (2) |
| H2A | 0.3501 | 0.5817 | -0.0364 | 0.105* |
| H2B | 0.3118 | 0.7328 | -0.0606 | 0.105* |
| C3 | 0.3086 (4) | 0.8759 (10) | 0.0530 (4) | 0.100 (3) |
| H3A | 0.2718 | 0.8149 | 0.0323 | 0.150* |
| H3B | 0.3142 | 0.9721 | 0.0305 | 0.150* |
| H3C | 0.2946 | 0.8959 | 0.0969 | 0.150* |
| C4 | 0.3862 (5) | 0.6584 (14) | -0.1161 (5) | 0.118 (4) |
| H4A | 0.4322 | 0.6047 | -0.1115 | 0.141* |
| H4B | 0.3965 | 0.7587 | -0.1349 | 0.141* |
| C5 | 0.4287 (5) | 0.8563 (10) | 0.0980 (4) | 0.097 (2) |
| H5A | 0.4704 | 0.7899 | 0.1009 | 0.145* |
| H5B | 0.4062 | 0.8644 | 0.1397 | 0.145* |
| H5C | 0.4436 | 0.9571 | 0.0838 | 0.145* |
| N2 | 0.3409 (3) | 0.5710 (7) | -0.1617 (2) | 0.0762 (14) |
| HN2A | 0.2954 | 0.6056 | -0.1598 | 0.114* |
| HN2B | 0.3418 | 0.4712 | -0.1513 | 0.114* |
| HN2C | 0.3581 | 0.5833 | -0.2015 | 0.114* |

## *Atomic displacement parameters (Å$^2$)*

| | $U^{11}$ | $U^{22}$ | $U^{33}$ | $U^{12}$ | $U^{13}$ | $U^{23}$ |
|---|---|---|---|---|---|---|
| Pb1 | 0.03485 (9) | 0.03864 (9) | 0.04655 (10) | 0.00044 (5) | -0.00120 (5) | 0.00026 (5) |
| I1 | 0.05734 (19) | 0.0740 (2) | 0.04621 (16) | 0.00896 (14) | 0.00173 (13) | 0.00028 (13) |
| I2 | 0.05219 (18) | 0.05186 (18) | 0.0748 (2) | -0.01901 (14) | -0.00041 (15 | -0.00699 (15) |
| I3 | 0.0643 (2) | 0.0923 (3) | 0.04806 (16) | -0.00127 (18) | -0.00170 (14 | -0.00504 (15) |
| I4 | 0.05421 (19) | 0.05191 (17) | 0.0760 (2) | 0.02034 (13) | 0.00764 (15) | 0.00426 (14) |
| N1 | 0.060 (3) | 0.058 (2) | 0.067 (3) | -0.0034 (18) | -0.0035 (19) | 0.003 (2) |
| C1 | 0.059 (3) | 0.113 (5) | 0.060 (3) | 0.019 (3) | -0.004 (2) | -0.018 (3) |
| C2 | 0.079 (4) | 0.114 (6) | 0.070 (4) | -0.015 (4) | -0.009 (3) | -0.001 (4) |
| C3 | 0.074 (5) | 0.132 (7) | 0.093 (5) | 0.030 (4) | 0.027 (4) | 0.029 (4) |
| C4 | 0.081 (5) | 0.153 (9) | 0.119 (7) | -0.013 (5) | -0.006 (5) | -0.066 (7) |
| C5 | 0.111 (7) | 0.104 (6) | 0.076 (4) | -0.023 (5) | -0.014 (4) | -0.017 (4) |
| N2 | 0.085 (4) | 0.079 (3) | 0.064 (3) | 0.009 (3) | -0.001 (2) | -0.019 (2) |

*Geometric parameters (Å, º)*

| | | | |
|---|---|---|---|
| Pb1—I1 | 3.1238 (4) | C2—C4 | 1.347 (12) |
| Pb1—I2 | 3.1565 (4) | C2—H2A | 0.9700 |
| Pb1—I4 | 3.1872 (5) | C2—H2B | 0.9700 |
| Pb1—I2$^{i}$ | 3.1895 (5) | C3—H3A | 0.9600 |
| Pb1—I4$^{ii}$ | 3.2578 (4) | C3—H3B | 0.9600 |
| Pb1—I3 | 3.3015 (4) | C3—H3C | 0.9600 |
| I2—Pb1$^{iii}$ | 3.1895 (5) | C4—N2 | 1.470 (9) |
| I4—Pb1$^{iv}$ | 3.2577 (4) | C4—H4A | 0.9700 |
| N1—C5 | 1.451 (9) | C4—H4B | 0.9700 |
| N1—C3 | 1.455 (9) | C5—H5A | 0.9600 |
| N1—C1 | 1.475 (7) | C5—H5B | 0.9600 |
| N1—HN1 | 0.8200 | C5—H5C | 0.9600 |
| C1—C2 | 1.556 (10) | N2—HN2A | 0.8900 |
| C1—H1A | 0.9700 | N2—HN2B | 0.8900 |
| C1—H1B | 0.9700 | N2—HN2C | 0.8900 |
| | | | |
| I1—Pb1—I2 | 93.335 (11) | C4—C2—H2A | 109.1 |
| I1—Pb1—I4 | 88.936 (10) | C1—C2—H2A | 109.1 |
| I2—Pb1—I4 | 93.404 (17) | C4—C2—H2B | 109.1 |
| I1—Pb1—I2$^{i}$ | 91.058 (11) | C1—C2—H2B | 109.1 |
| I2—Pb1—I2$^{i}$ | 86.532 (14) | H2A—C2—H2B | 107.8 |
| I4—Pb1—I2$^{i}$ | 179.935 (13) | N1—C3—H3A | 109.5 |
| I1—Pb1—I4$^{ii}$ | 91.492 (11) | N1—C3—H3B | 109.5 |
| I2—Pb1—I4$^{ii}$ | 174.941 (12) | H3A—C3—H3B | 109.5 |
| I4—Pb1—I4$^{ii}$ | 85.165 (14) | N1—C3—H3C | 109.5 |
| I2$^{i}$—Pb1—I4$^{ii}$ | 94.899 (16) | H3A—C3—H3C | 109.5 |
| I1—Pb1—I3 | 174.858 (12) | H3B—C3—H3C | 109.5 |
| I2—Pb1—I3 | 91.494 (11) | C2—C4—N2 | 116.5 (8) |
| I4—Pb1—I3 | 92.544 (11) | C2—C4—H4A | 108.2 |
| I2$^{i}$—Pb1—I3 | 87.468 (11) | N2—C4—H4A | 108.2 |
| I4$^{ii}$—Pb1—I3 | 83.731 (11) | C2—C4—H4B | 108.2 |
| Pb1—I2—Pb1$^{iii}$ | 167.679 (15) | N2—C4—H4B | 108.2 |
| Pb1—I4—Pb1$^{iv}$ | 163.126 (15) | H4A—C4—H4B | 107.3 |
| C5—N1—C3 | 111.7 (7) | N1—C5—H5A | 109.5 |
| C5—N1—C1 | 111.5 (5) | N1—C5—H5B | 109.5 |
| C3—N1—C1 | 113.0 (5) | H5A—C5—H5B | 109.5 |
| C5—N1—HN1 | 109.5 | N1—C5—H5C | 109.5 |
| C3—N1—HN1 | 110.1 | H5A—C5—H5C | 109.5 |
| C1—N1—HN1 | 100.5 | H5B—C5—H5C | 109.5 |
| N1—C1—C2 | 109.7 (6) | C4—N2—HN2A | 109.5 |
| N1—C1—H1A | 109.7 | C4—N2—HN2B | 109.5 |
| C2—C1—H1A | 109.7 | HN2A—N2—HN2B | 109.5 |
| N1—C1—H1B | 109.7 | C4—N2—HN2C | 109.5 |
| C2—C1—H1B | 109.7 | HN2A—N2—HN2C | 109.5 |
| H1A—C1—H1B | 108.2 | HN2B—N2—HN2C | 109.5 |
| C4—C2—C1 | 112.7 (7) | | |
| | | | |
| C5—N1—C1—C2 | 168.7 (7) | N1—C1—C2—C4 | -176.8 (9) |
| C3—N1—C1—C2 | -64.5 (8) | C1—C2—C4—N2 | -177.9 (8) |

Symmetry codes: (i) -*x*+1, *y*-1/2, -*z*+1/2; (ii) -*x*+1/2, *y*-1/2, *z*; (iii) -*x*+1, *y*+1/2, -*z*+1/2; (iv) -*x*+1/2, *y*+1/2, *z*.

## **Table S7.** DMPA (1): Hydrogen-bond geometry (Å,°)

| *D*—H···*A* | *D*—H | H···*A* | *D*···*A* | *D*—H···*A* |
|---|---|---|---|---|
| N1—H*N*1···I1 | 0.82 | 2.88 | 3.665 (5) | 160 |
| C1—H1*A*···I1$^{v}$ | 0.97 | 3.17 | 4.035 (6) | 149 |
| C1—H1*B*···I3$^{vi}$ | 0.97 | 3.24 | 4.101 (7) | 148 |
| C2—H2*A*···I1 | 0.97 | 3.14 | 3.950 (8) | 142 |
| C2—H2*B*···I3$^{vii}$ | 0.97 | 3.16 | 3.940 (8) | 139 |
| C3—H3*C*···I4 | 0.96 | 3.31 | 3.988 (8) | 129 |
| C4—H4*A*···I1$^{v}$ | 0.97 | 3.17 | 4.030 (9) | 148 |
| C5—H5*B*···I2$^{iii}$ | 0.96 | 3.30 | 3.814 (7) | 116 |
| N2—H*N*2*A*···I3$^{vii}$ | 0.89 | 2.81 | 3.665 (6) | 161 |
| N2—H*N*2*B*···I3$^{viii}$ | 0.89 | 2.77 | 3.655 (6) | 178 |
| N2—H*N*2*C*···I2$^{v}$ | 0.89 | 3.07 | 3.644 (5) | 124 |

Symmetry codes: (iii) -*x*+1, *y*+1/2, -*z*+1/2; (v) -*x*+1, -*y*+1, -*z*; (vi) *x*, -*y*+3/2, *z*-1/2; (vii) -*x*+1/2, -*y*+1, *z*-1/2; (viii) *x*, -*y*+1/2, *z*-1/2.

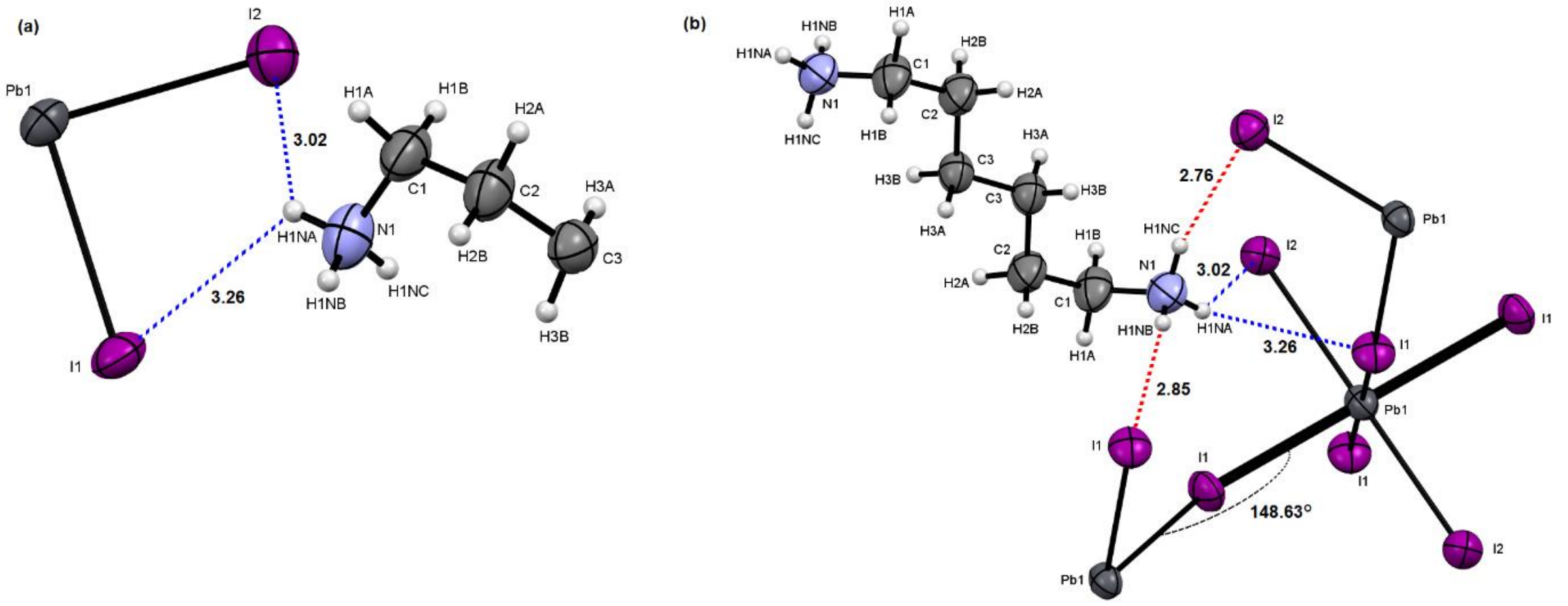

**Figure S3**. HE: (a) A view of the asymmetric unit with the atomic labelling scheme. (b) A view of the local environment of the asymmetric unit showing the distortion of the in-plane inorganic layer [the Pb1–I1–Pb1 angle is 148.63(2)°, see Table S2]. The broken blue lines indicate two of the N—H···I hydrogen bonds listed in Table S3 (*i.e.*, hydrogen-acceptor distances 3.26 and 3.02 Å), involving one hydrogen atom and two iodine atoms of the asymmetric unit. The broken red lines relate hydrogen atoms of the asymmetric unit and symmetry equivalent iodine atoms (*i.e.*, hydrogen-acceptor distances 2.85 and 2.76 Å, see Table S3). Ellipsoids are drawn at 50% of probability level.

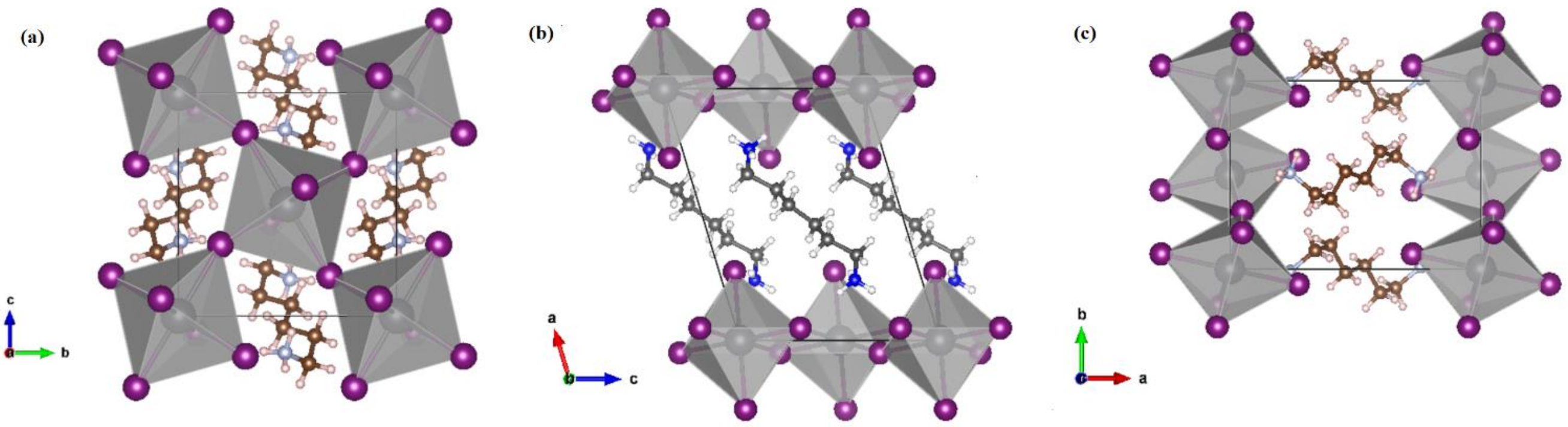

**Figure S4.** HE: A view along *a* (a), along *b* (b) and along *c* (c) of the crystal packing.

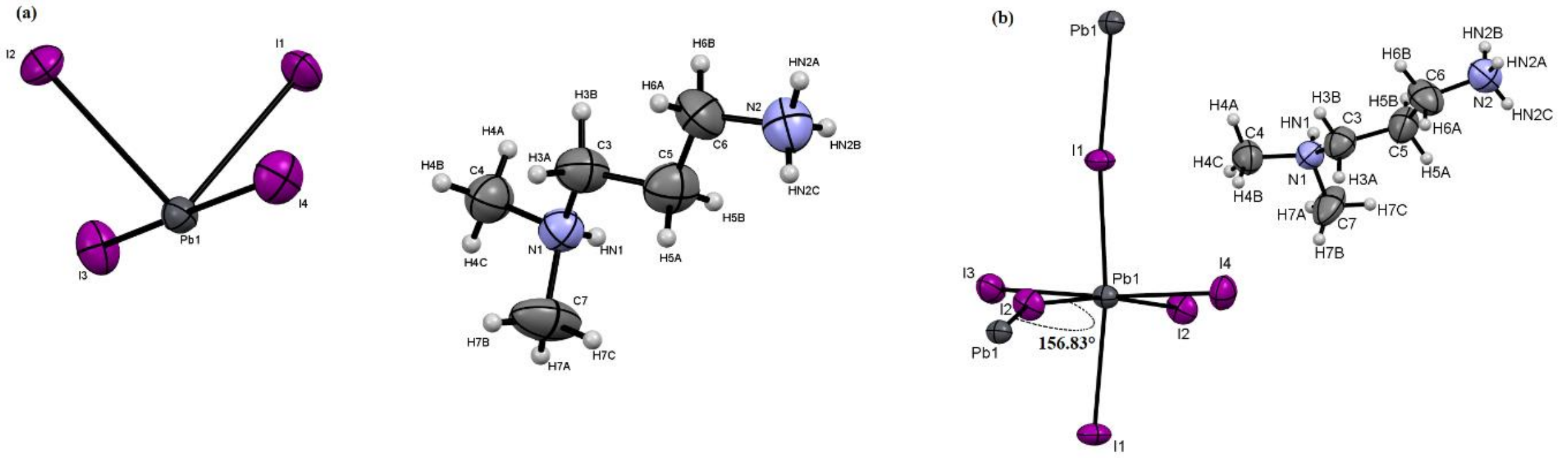

**Figure S5.** DMPA (2): (a) A view of the asymmetric unit with the atomic labelling scheme. (b) A view of the local environment of the asymmetric unit showing the distortion of the in-plane inorganic layer [the Pb1–I2–Pb1 angle is 156.83°, see Table S4]. Ellipsoids are drawn at 50% of probability level.

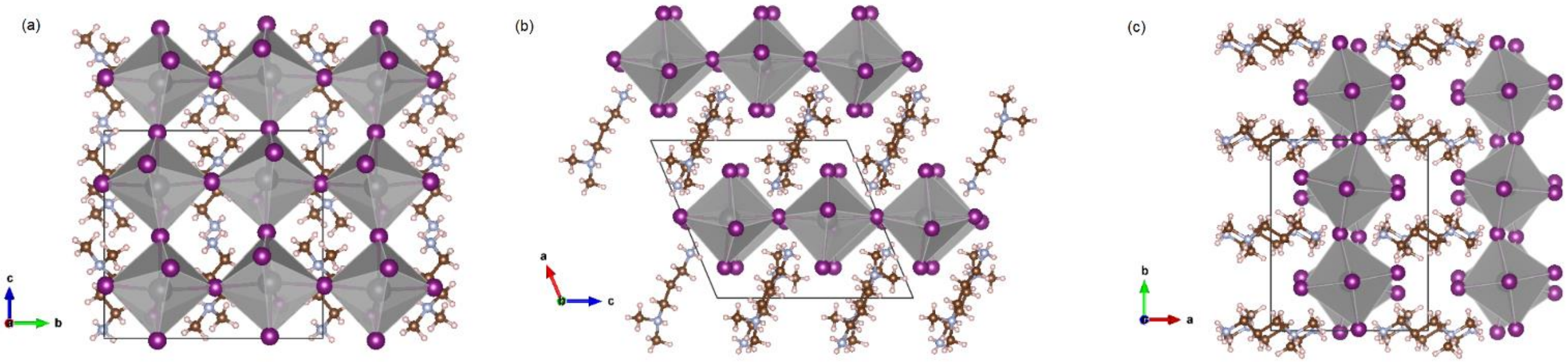

**Figure S6.** DMPA (2): A view along *a* (a), along *b* (b) and along *c* (c) of the crystal packing.

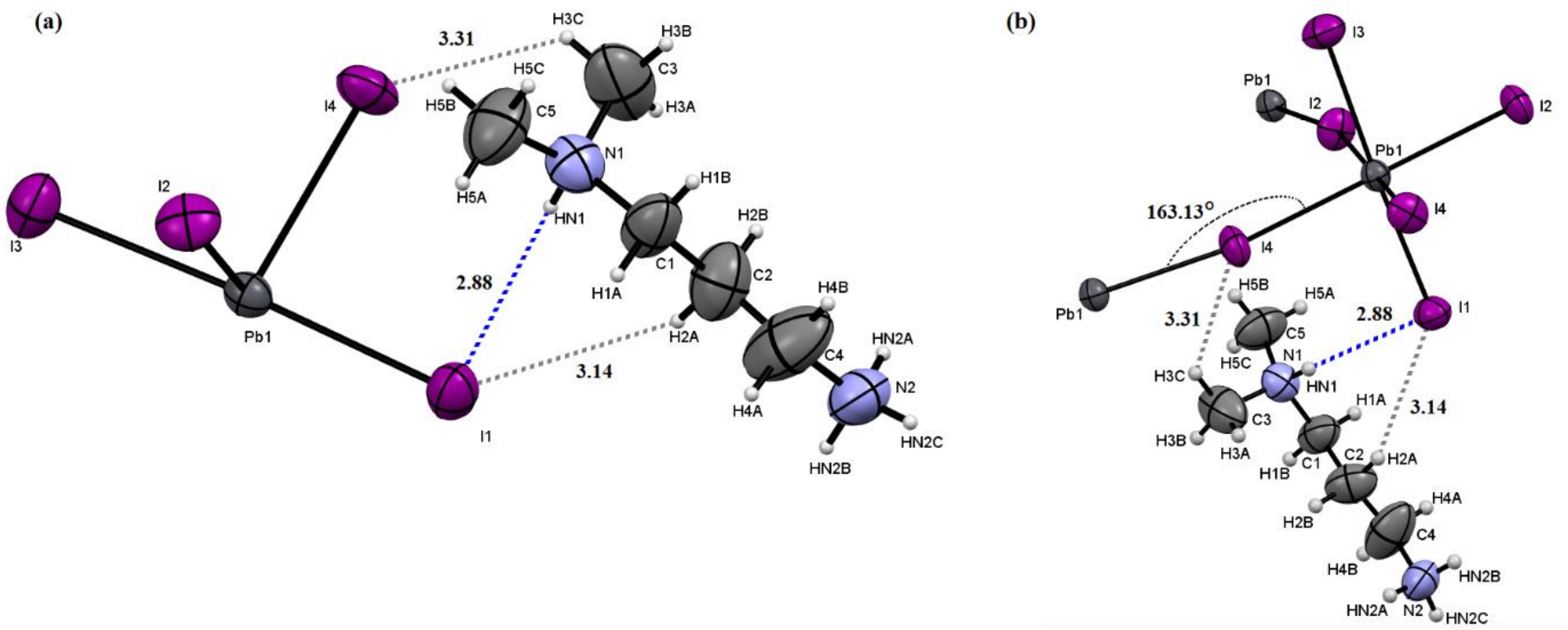

**Figure S7.** DMPA (1): (a) A view of the asymmetric unit with the atomic labelling scheme. (b) A view of the local environment of the asymmetric unit showing the distortion of the in-plane inorganic layer [the Pb1–I4–Pb1 angle is 163.13°, see Table S6]. The broken blue lines indicate the N1—HN1···I1 hydrogen bond listed in Table S7 (*i.e.*, hydrogen-acceptor distance 2.88 Å), involving one hydrogen atom and one iodine atom of the asymmetric unit. The broken dark grey lines relate hydrogen (bonded to a carbon atom) and iodine atoms of the asymmetric unit (*i.e.*, hydrogen-acceptor distances 3.14 and 3.31 Å, see Table S7). Ellipsoids are drawn at 50% of probability level.

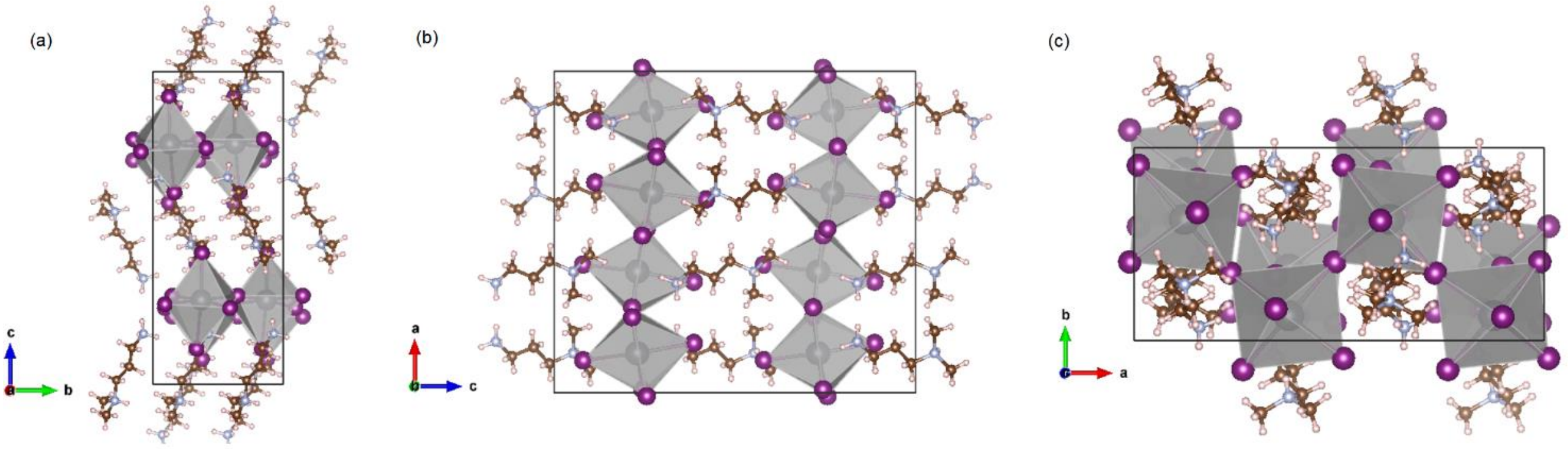

**Figure S8.** DMPA (1): A view along *a* (a), along *b* (b) and along *c* (c) of the crystal packing.

*Penetration depth*

Table S8 shows $NH_3$ penetration in terms of average distance of nitrogen atoms for each compound studied in this work. To properly argue the penetration depth, we reported also the value estimated for PEAI-F perovskite single crystals (Figure S13).

| Perovskite | Pb-I-Pb (°) | **$<d_N>_{U\&L}$** (Å) |
|---|---|---|
| DMPA (1) | 163.13° | 1.0565 |
| DMPA (2) | 156.83° | 0.9315 |
| PEAI-F | 151.77° | 0.576 |
| HE | 148.63° | 0.373 |

**Table S8.** DMPA (1), DMPA (2), PEAIF and HE: in-plane distortion angle ($Pb_{eq}$–I–$Pb_{eq}$) and average distance between the nitrogen atoms belonging to the terminal $NH_3$ group and the nearest least-squares planes through the upper and lower axial iodine atoms of the $PbI_6$ octahedra ($<d_N>_{U\&L}$).

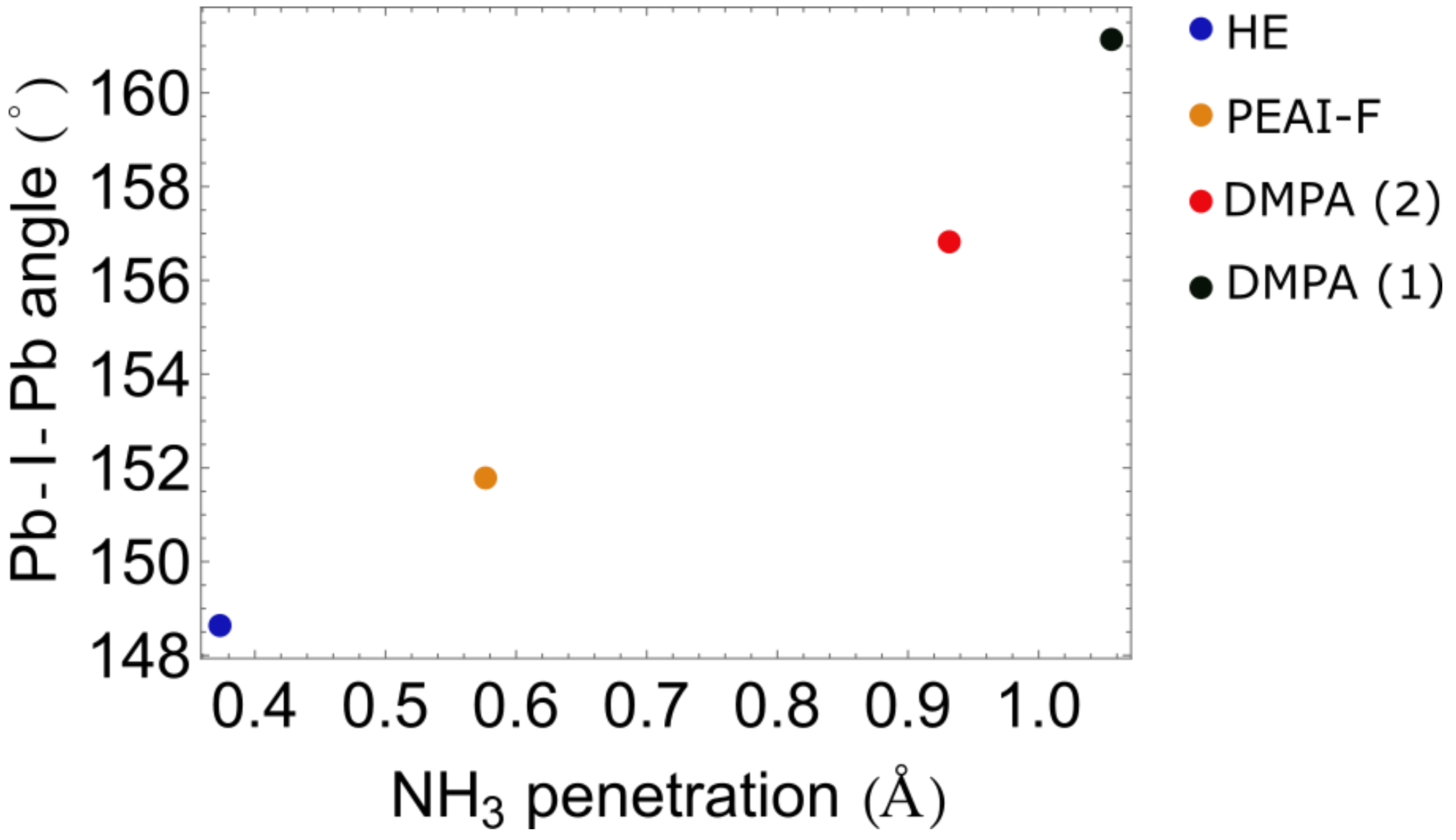

**Figure S9**. DMPA (1), DMPA (2), PEAI-F and HE: $Pb_{eq}$–I–$Pb_{eq}$ bond angle as a function of the organic cation penetration.

In Figure S10, S11, S12 and S13 a view of the inorganic chain, the two nearest organic layers and the two least-squares planes through the upper and lower axial iodine atoms of the $PbI_6$ octahedra (*i.e.*, $plane_U$ and $plane_L$, the pink planes in Figure S10, S11, S12 and S13) is given in the case of DMPA(1), DMPA(2), HE and PEAI-F, respectively.

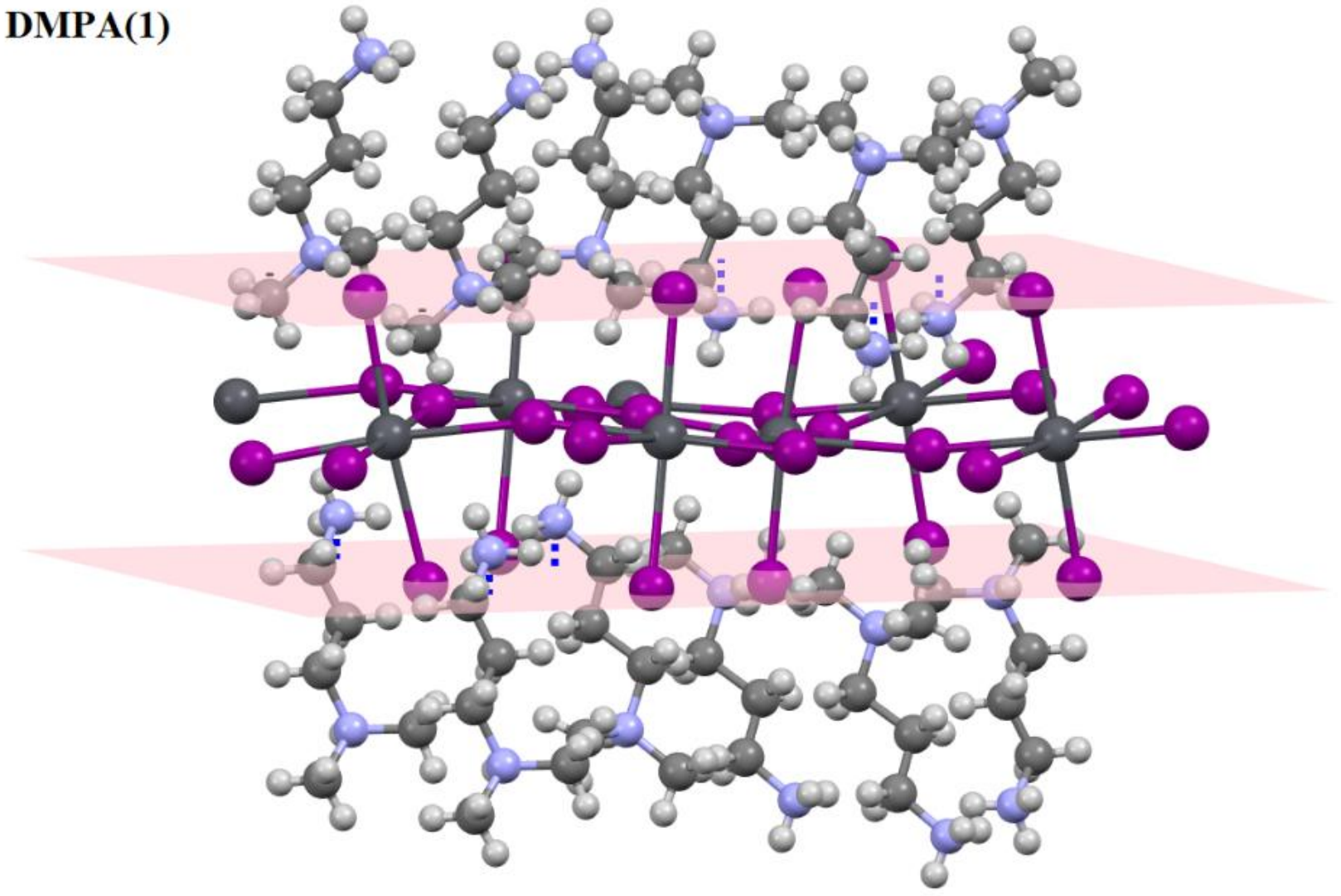

**Figure S10.** DMPA (1): a view of the inorganic chain, the two nearest organic layers and the two least-squares planes through the upper and lower axial iodine atoms of the $PbI_6$ octahedra (pink planes). The distances between the N atoms of the terminal $NH_3$ group and the two least-squares planes are drawn as broken blue lines.

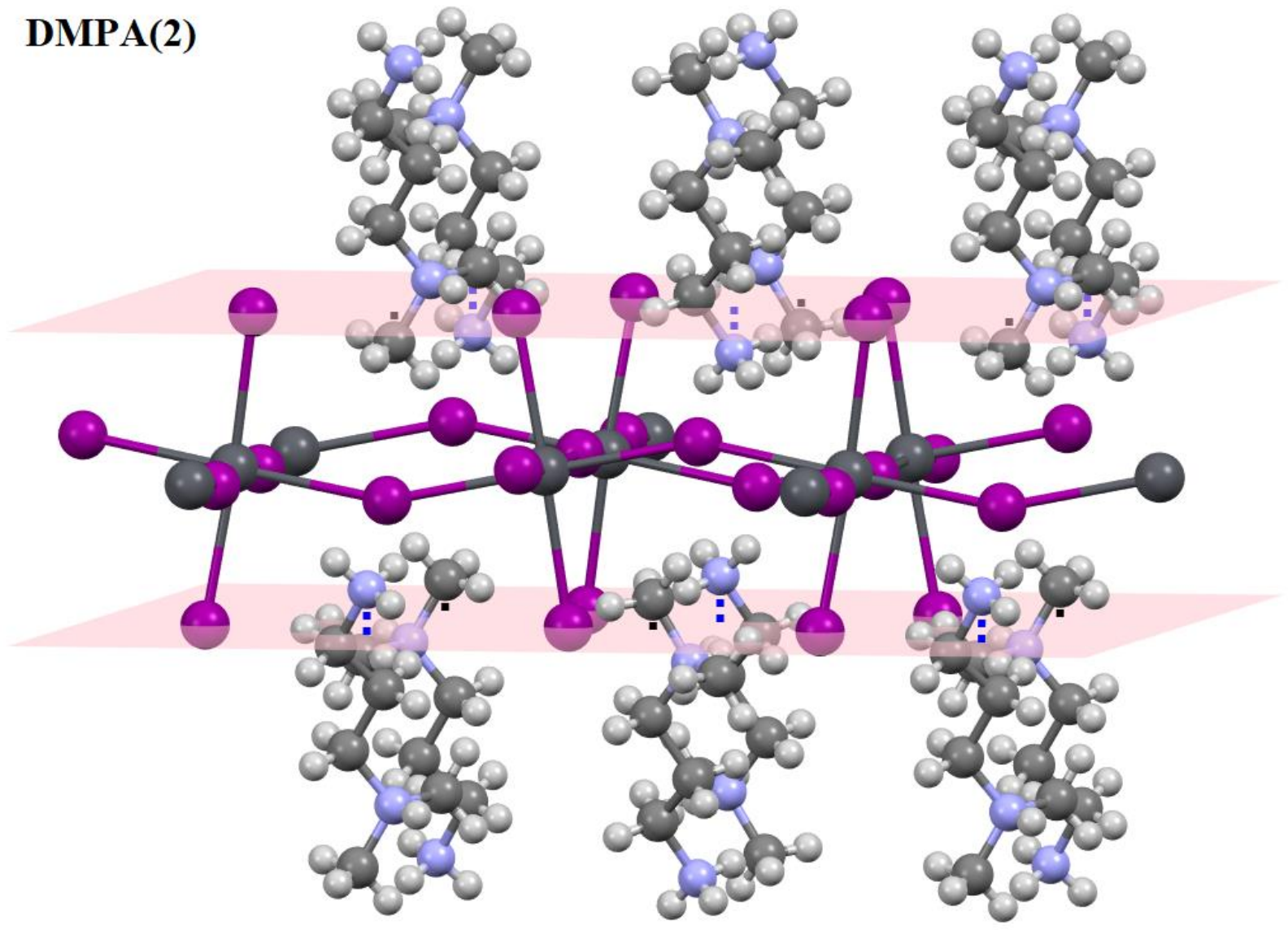

**Figure S11.** DMPA (2): a view of the inorganic chain, the two nearest organic layers and the two least-squares planes through the upper and lower axial iodine atoms of the $PbI_6$ octahedra (pink planes). The distances between the terminal N atoms of the $NH_3$ group and the two least-squares planes are drawn as broken blue lines.

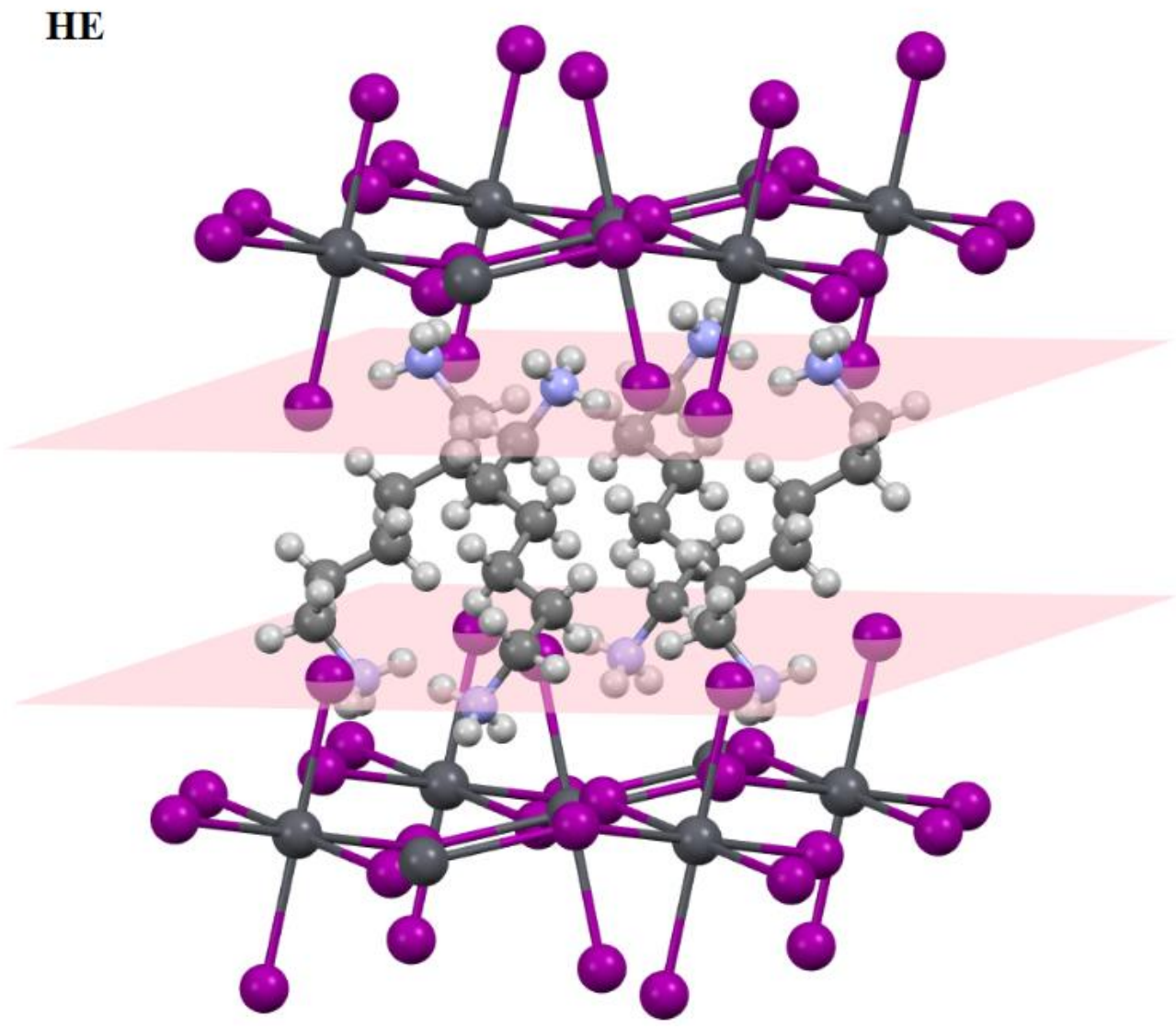

**Figure S12.** HE: a view of the inorganic chain, the two nearest organic layers and the two least-squares planes through the upper and lower axial iodine atoms of the $PbI_6$ octahedra (pink planes).

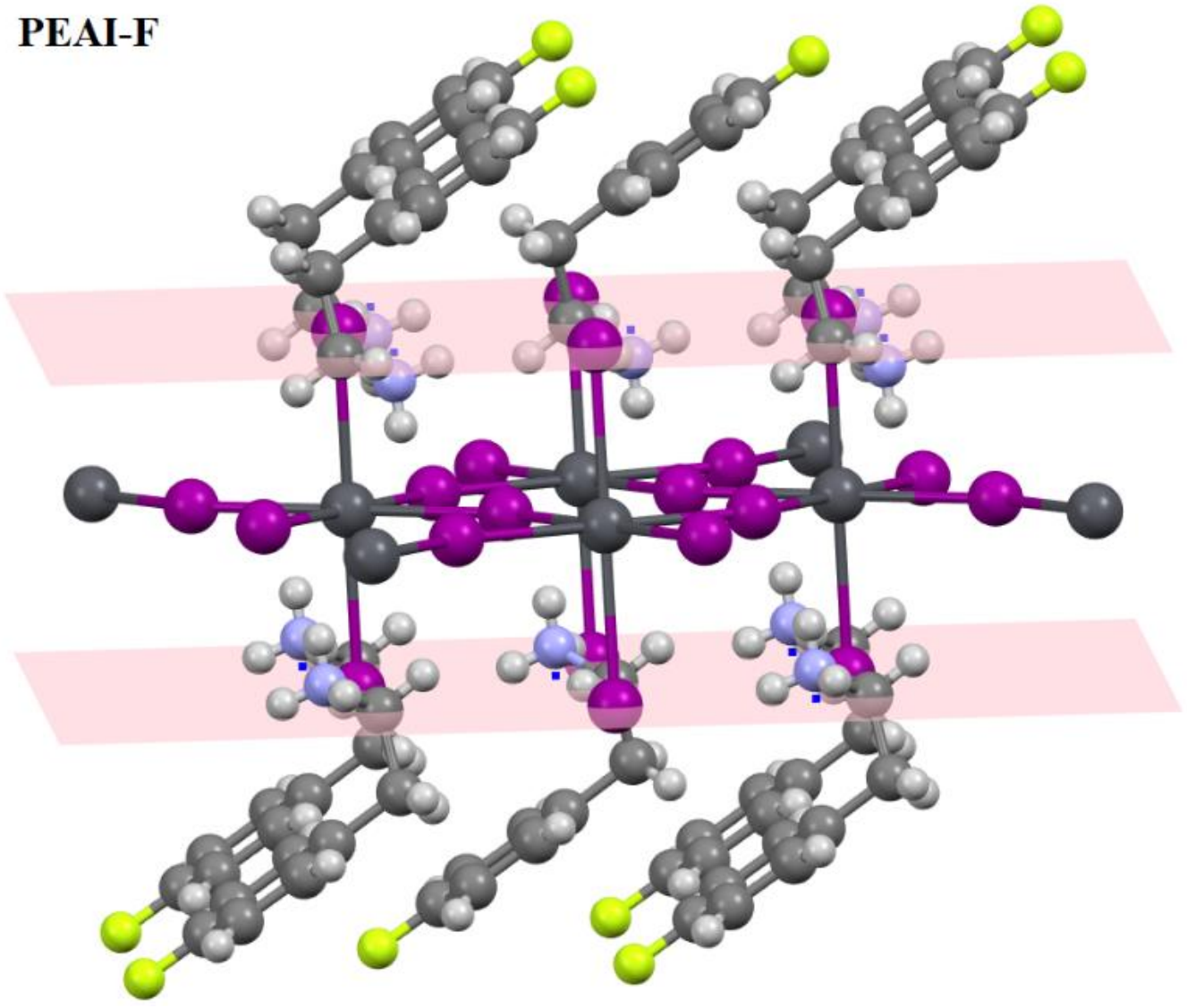

**Figure S13.** PEAI-F: a view of the inorganic chain, the two nearest organic layers and the two least-squares planes through the upper and lower axial iodine atoms of the $PbI_6$ octahedra (pink planes). The distances between the N atoms of the terminal $NH_3$ group and the two least-squares planes are drawn as broken blue lines.

## Optical measurements

Photoluminescence measurements at room temperature were performed in reflection configuration with a continuous-wave laser, centered at $\lambda$ = 405 nm, by using a home-built microscope, collecting the signal using a 40x objective (NA 0.95), while the absorption spectra were collected exciting the perovskite flakes with a Xenon white lamp, in transmission configuration. A 75 cm lens was additionally used to perform Fourier spectroscopy. A linear polarizer and a half-wave plate placed in front of the spectrometer allow to resolve the TE and TM polarizations. Temperature-dependent photoluminescence measurements were performed by using a nitrogen-filled cryostat, by exciting the sample with a continuous-wave laser, centered at $\lambda$ = 405 nm, collecting the signal using a long working distance 40x objective (NA 0.6). The signal was detected with a 300 mm spectrometer (Acton Spectra ProSP-2300, Princeton Instruments, USA) coupled to a Charge Coupled Device (CCD Pixis eXcelon 400, Princeton Instruments, USA). The spectrometer was equipped with two gratings, 300 g/mm and 1200 g/mm, both of them blazed at 500 nm.

Figure S14 shows PL and Abs spectra for HE perovskite (Figure S14a, Blue continuous line and black dashed line, respectively) and for DMPA (2) perovskite (Figure S14b, Red continuous line and black dashed line, respectively).

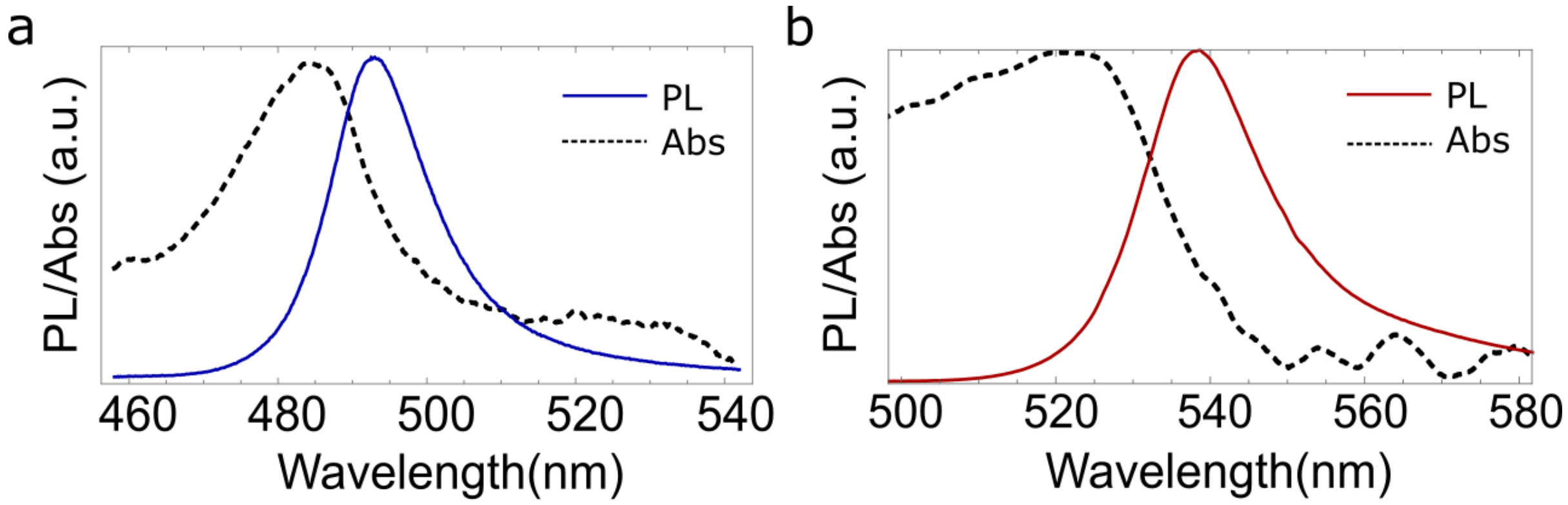

**Figure S14.** (a) PL (Blue, continuous line) and Abs (Black, dashed line) spectra of HE single crystals on glass substrate. (b) PL (Red, continuous line) and Abs (Black, dashed line) spectra of DMPA (2) single crystals on glass substrate.

## Temperature-dependent PL

Figure S15a and S15b show temperature-dependent photoluminescence of HE and stable DMPA (2) phase perovskites from 77K to RT.

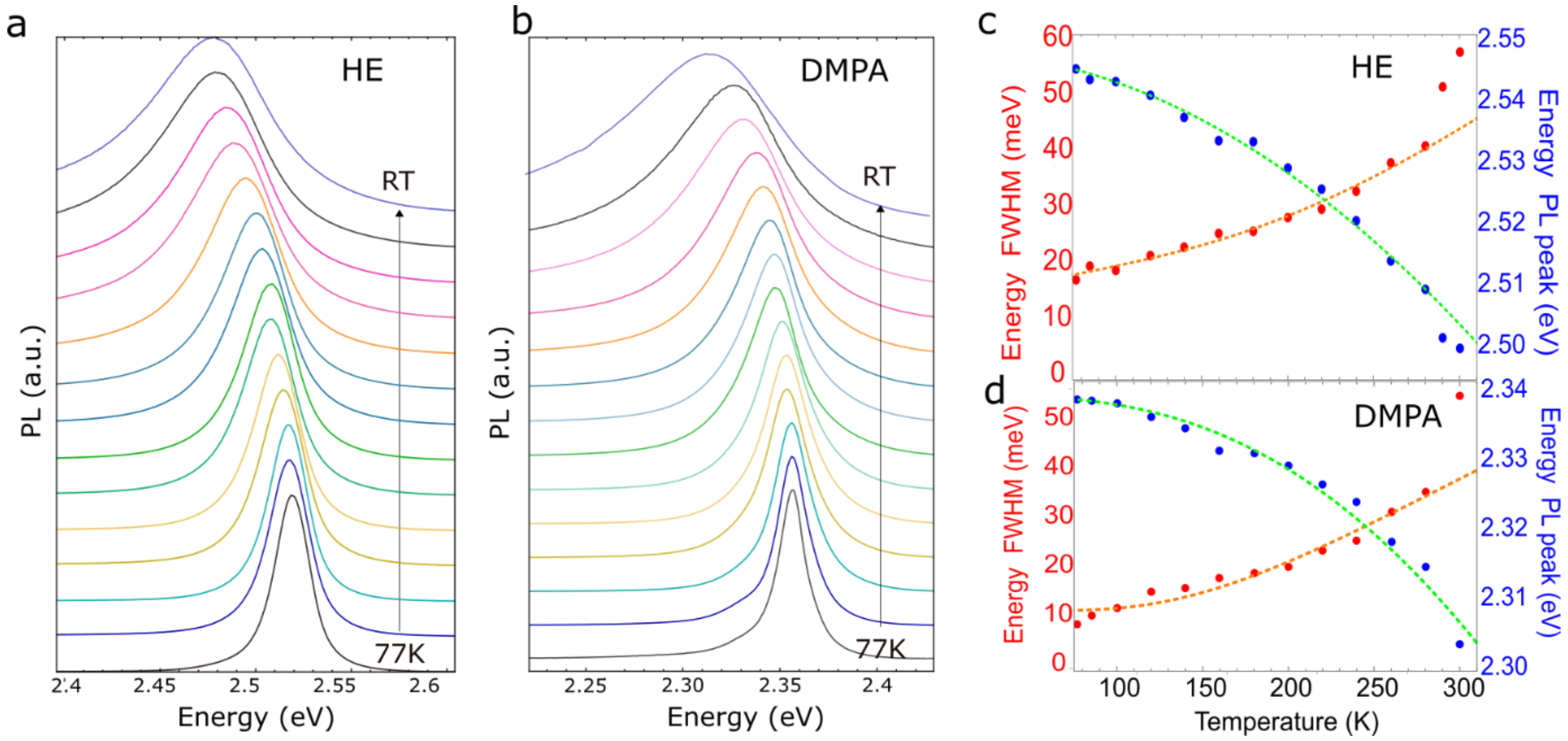

**Figure S15.** Temperature-dependent PL spectra of HE (a) and DMPA (2) (b) single crystal on glass substrate. The spectra are vertically spaced for clarity. Temperature dependence of the maximum of the photoluminescence peak (blue dots) and of FWHM (red dots) for HE-based perovskite (c) and DMPA (2) (d)-based perovskite (d). Green and orange dashed line are obtained by fitting the experimental data with the Varshni and Boson equation respectively.

The temperature dependence of PL energy peak and FWHM for the two perovskites is very similar. Figure S15c and S15d report the dependence of the energy position of the excitonic resonance ($E(T)$) for HE (Figure S15c) and DMPA (2) (Figure S15d) marked as blue dots. Green dashed lines are obtained by fitting the experimental data with the Varshni equation[17] (1):

$$E(T) = E_0 - \frac{\alpha T^2}{T + \beta} \quad (1)$$

Where $E_0$ is the energy gap at 0K, α is a constant and β is related to the Debye temperature. The extracted parameters are reported in Table S9.

| **Perovskite** | $\boldsymbol{E_0}$ **(eV)** | **α (eV/K)** | **β (K)** |
|---|---|---|---|
| **HE** | 2.551 | $2.2\ x\ 10^{-4}$ | 156.52 |
| **DMPA (2)** | 2.348 | $2.02\ x\ 10^{-4}$ | 177.34 |

**Table S9.** Extracted parameters from the Varshni model for HE and DMPA (2) perovskites.

Moreover, by increasing the temperature the PL peak broadening for both HE and DMPA (2) perovskites is clearly measurable, analysing the FWHM trend as function of the temperature ($\Gamma(T)$) (Red dots in Figure S15c and S15d). The experimental data are fitted with the Boson model (2) (orange dashed line):

$$\Gamma(T) = \Gamma_0 + \Gamma_{LO} * exp\left(\frac{E_{LO}}{k_b T} - 1\right)^{-1} \tag{2}$$

Where $\Gamma_0$ is the inhomogeneous broadening contribution, that is not dependent from the temperature and is caused by the inhomogeneity of the sample; $\Gamma_{LO}$ the excitonic longitudinal-mode optical phonon coupling coefficient and $E_{LO}$ is the longitudinal-mode optical phonon energy.

| **Perovskite** | **$\Gamma_0$ (meV)** | **$\Gamma_{LO}$ (meV)** | **$E_{LO}$(meV)** |
|---|---|---|---|
| **HE** | 12.997 | 199.09 | 73.4 |
| **DMPA (2)** | 11.484 | 191.03 | 54.05 |

**Table S10.** Extracted parameters from the Boson model for HE and DMPA (2) perovskites.

The broadening of the PL peak is attributed to the phonon assisted broadening effect. From the fit of the experimental data we are able to extract the energy of phonon modes ($E_{LO}$) that play a crucial role in electron–phonon coupling.[18] The extracted parameters, reported in Table S10, are similar to other reported in previous work for classical RP perovskites.[19] The $\Gamma_0$ and $\Gamma_{LO}$ values are smaller for DMPA (2) than HE, indicating that DMPA (2) structure is more ordered, as observed from XRD characterization, and the non-radiative loss are smaller. This can be explained by considering that the longer alkyl–ammonium chain (C6) promotes the exciton–phonon scattering, as observed from Zhang et al.[19]

## Template confined growth

*PDMS template fabrication*

The patterned elastomeric template, realized by a conventional soft-lithography technique, was obtained starting from a silicon master (ThunderNIL) having an array of microchannels with a height (h) of 300 nm and a width (w) of 6 µm.

PDMS (Sylgard 184, Dow-Corning) was obtained by mixing base and curing agent in a 9:1 ratio and kept it in a desiccator for 30 minutes. After all air bubbles have been carefully removed from PDMS, the liquid elastomer was casted on the silicon master. The system was then cured in an oven for 15 minutes at 140°C. Finally, the elastomeric replica is peeled off from the master thus obtaining a patterned microfluidic device.

*Precursor solution*

An amount of HE and DMPA (2) (290.8 mg and 286.6 mg respectively) crystals obtained from solution-based growth method and washed with DE were redissolved in GBL (1 ml) in order to obtain a 0.35 M solution, under stirring and heating at 90°C.

*HE and DMPA (2) microwire growth*

A PDMS mold with the pattern, having desired features and dimensions, is placed in conformal contact on a glass substrate, previously washed with acetone and IPA in an ultrasonic bath. In this way a network of closed micro-channels is obtained and within which the perovskite precursor solution can flow. Then, 1 µl of precursors solution was dropped at one end of the capillaries. The solution, driven by the capillary force, soaks the PDMS microchannels and fills them completely. The microfluidic device, fulfilled with precursor solution, was then kept in a transparent box and opportunely closed with parafilm, in order to enhance the pressure within microchannels and obtain a supersaturation state at room temperature. After 10 hours, when the solvent is completely evaporated, the PDMS template was removed from the substrate so as to obtain millimeter-long perovskites microwires and lateral dimensions and heights precisely defined by PDMS mold. Fig. S16 shows a SEM image of the as- grown HE single crystals on glass substrate.

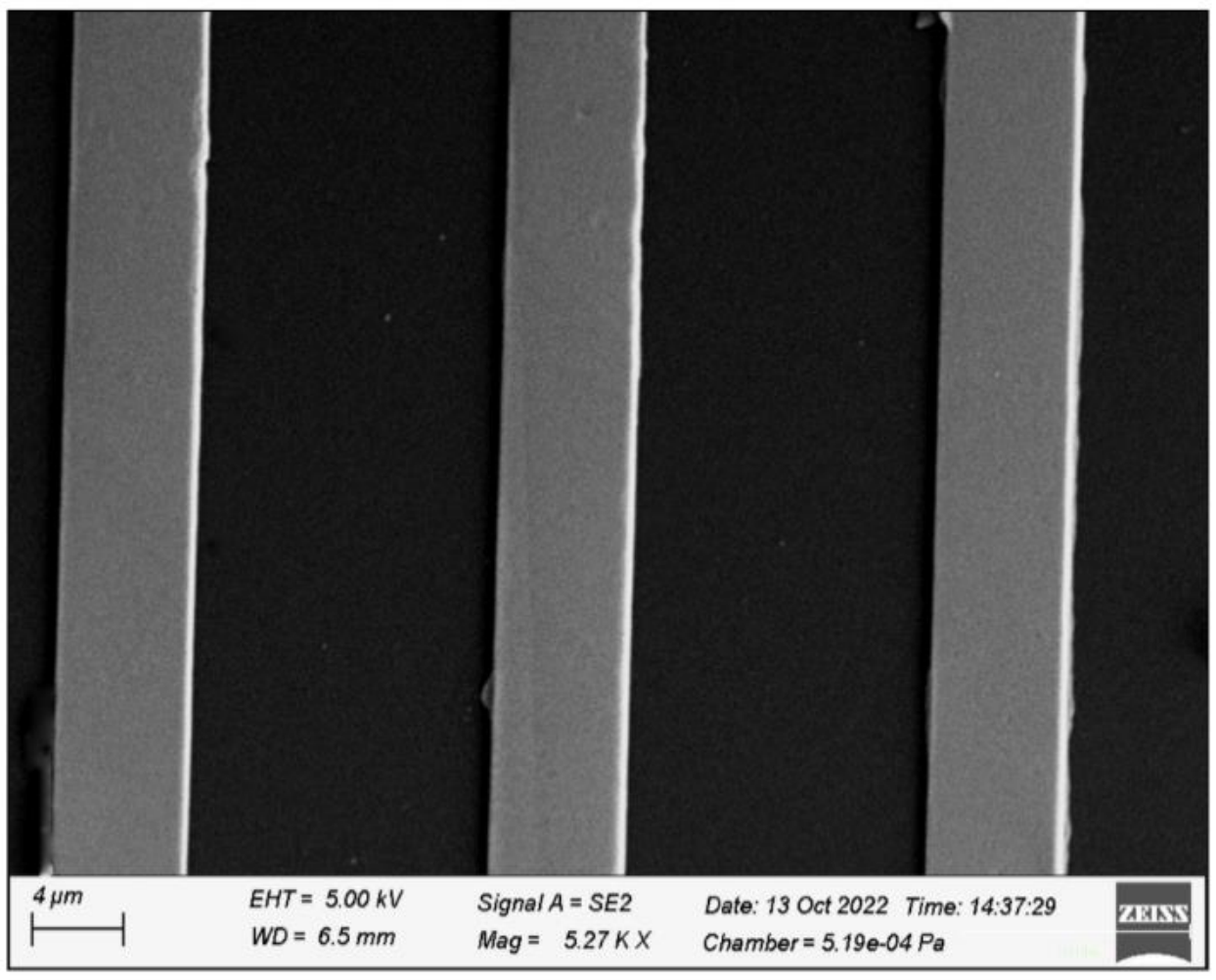

**Figure S16.** SEM image of HE perovskite microwires grown on glass substrate.

**Electron – beam lithography (EBL)**

The grating consists in an array of stripes, with a pitch of 300 nm and a filling factor (FF) of 0.5; these parameters were obtained by simulating the system with S4 package[20] to extract the propagating mode in a 300nm-thick perovskite, considering a background refractive index of 1.8.

Electron-beam lithography was performed with a Raith system (VOYAGER) at an acceleration voltage of 30 KV and with a 26 pA current. The glass substrate with the grown microwires was first washed with 2-propanol, then a thin layer of polymethil methacrylate (PMMA A4, MicroChem) was spin coated at 4000 rpm for 60 sec, followed by soft baking at 180°C for 120 sec, obtaining a 180 nm thick resist layer. The sample was then cooled down to room temperature, gently washed in 2-propanol and dried with a nitrogen flow. A conductive resist was finally spin coated (DisCharge $H_2O$, MicroResist) at 1500rpm for 60 sec, and finally soft baked at 90°C for 60 sec. The sample was patterned at a 200 uC/cm$^2$ area dose, developed in a MIBK:2-propanol (3:1) mixture for 90 sec and finally washed in 2-propanol.

**Theoretical calculations**

To extract the Rabi splitting values from photoluminescence spectra reported in Figure 4c and 4d of the main text, we diagonalize a two coupled oscillator system matrix, described as follow:

$$\begin{pmatrix} \omega 0 + \omega C * (k) & \Omega R \\ \Omega R & Ex - i\gamma \end{pmatrix} \qquad (4)$$

Figure S17a and b show fitted dispersions in TE polarization for HE (Figure S17a) and DMPA (2) (Figure S17b) perovskite with a PMMA grating on top. The Rabi splitting for HE is 100 meV, while for DMPA (2) is 125 meV.

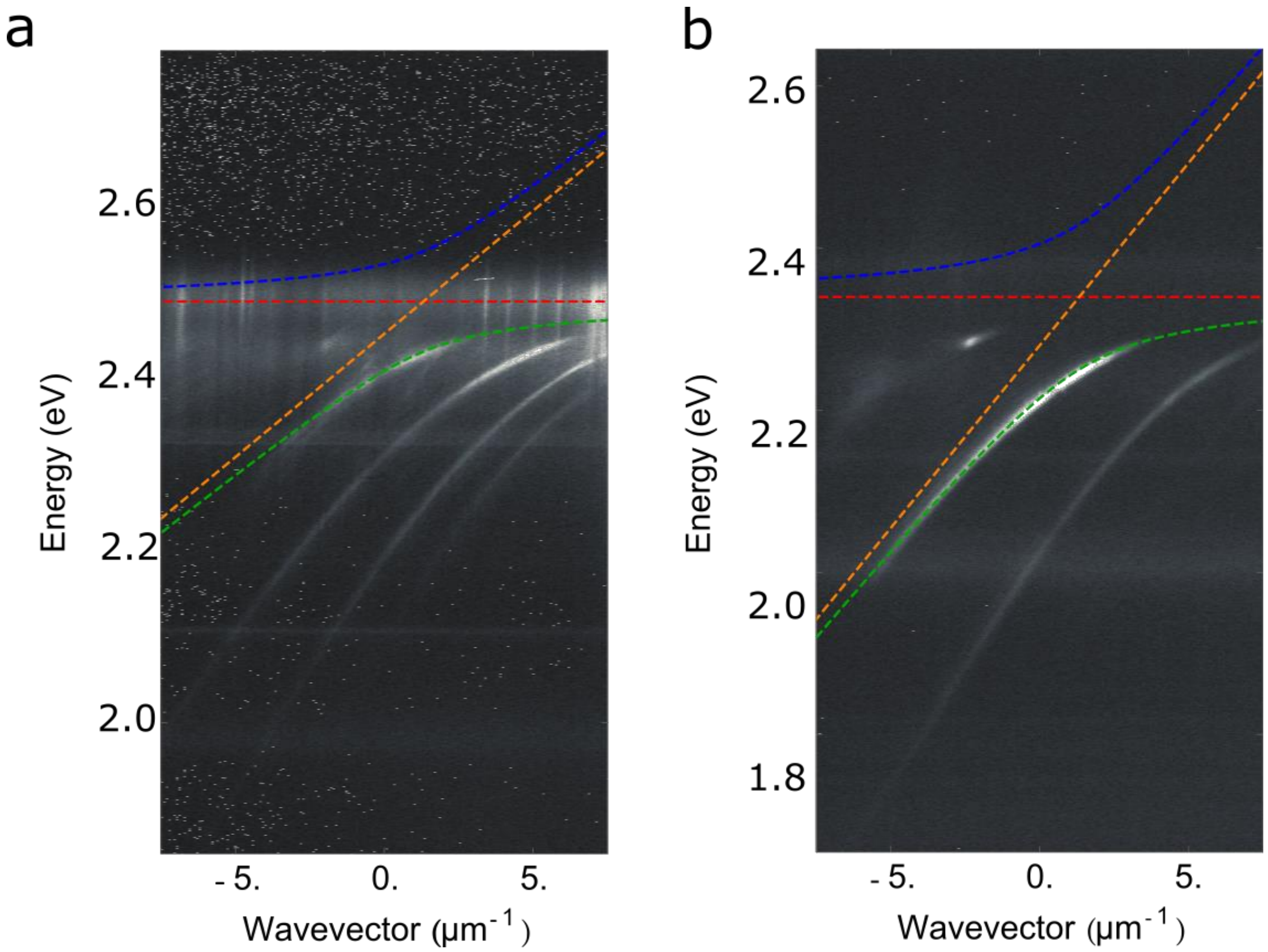

**Figure S17.** Fitted dispersions in TE polarization for HE (a) and DMPA (2) (b).